# Facile On-Chip Growth and Spatially Selective Color Tuning of $CsPbBr_3$@TFA Nanoplatelets via Ion Engineering

Sana Khan[1,2], Saeed Goudarzi[2], Lars Sonneveld[3], Felix Thiel[4], Stephan Schäffer[4], Bart Macco[5], Ke Ran[1,6,7], Sercan Ozen[8], Naho Kurahashi[9,10], Peter Haring Bolívar[4], Felix Lang[8], Bruno Ehrler[3], Burkay Uzlu[1], Thomas Riedl[9,10], Gerhard Müller-Newen[11], Surendra. B. Anantharaman[1,12], Maryam Mohammadi[1, *], Max. C. Lemme[1,2, *]

[1]*AMO GmbH, Otto-Blumenthal-Straße 25, 52074 Aachen, Germany*

[2]*RWTH Aachen University, Chair of Electronic Devices, Otto-Blumenthal-Str. 25, 52074 Aachen, Germany*

[3]*LMPV-Sustainable Energy Materials Department, AMOLF Science Park 104, 1098 XG Amsterdam, The Netherlands*

[4]*Institute for High Frequency and Quantum Electronics, University of Siegen, 57076 Siegen, Germany*

[5]*Department of Applied Physics and Science Education, Eindhoven University of Technology, P.O. Box 513, 5600, MB, Eindhoven, the Netherlands*

[6]*Central Facility for Electron Microscopy (GFE), RWTH Aachen University, Ahornstr. 55, 52074 Aachen, Germany*

[7]*Ernst Ruska Centre for Microscopy and Spectroscopy with Electrons ER-C, Forschungszentrum Jülich GmbH, Jülich 52428, Germany*

[8] *ROSI Freigeist Junior group, Soft Matter Physics and Optoelectronics, University of Potsdam, Potsdam, 14476, Germany*

[9]*Institute of Electronic Devices, University of Wuppertal, Rainer-Gruenter-Str. 21, 42119 Wuppertal, Germany*

[10]*Wuppertal Center for Smart Materials & Systems (CM@S), University of Wuppertal, Rainer-Gruenter-Str. 21, 42119 Wuppertal, Germany*

[11]*Institute of Biochemistry and Molecular Biology, Uniklinik RWTH Aachen, Pauwelsstrasse 30, Aachen, Germany*

[12]*Low-dimensional Semiconductors Lab, Department of Metallurgical and Materials Engineering, Indian Institute of Technology Madras, Chennai 600 036, India*

**Corresponding Authors:* mohammadi@amo.de; max.lemme@eld.rwth-aachen.de

**Abstract**

Metal halide perovskite (MHP) nanoplatelets are attractive for integrated optoelectronics because of their strong light-matter interactions and composition-tunable optical response. However, their high surface-to-volume ratio promotes surface defects and trap-assisted nonradiative recombination. In addition, their ionic nature facilitates rapid halide migration and limits selective composition tuning. Here, we combine in-synthesis (*in situ*) and post-synthesis ion engineering to suppress nonradiative recombination and enable selective on-chip color conversion. $CsPbBr_3$ nanoplatelets were grown directly on-chip via a solution process incorporating trifluoroacetate ($TFA^-$) pseudohalides. $CsPbBr_3$@TFA emits at 520 nm with a linewidth of 14.56 ± 0.5 nm and an average lifetime of 26.38 ns. X-ray photoelectron spectroscopy shows that $TFA^-$ passivates uncoordinated $Pb^{2+}$ and suppresses metallic lead ($Pb^0$) formation. Spatially selective halide exchange was then achieved using a photolithographically patterned PMMA/$Al_2O_3$ mask to confine HCl vapor exposure. In the unmasked regions, the photoluminescence peak shifts from 520 nm to 413 nm, producing laterally defined green- and violet-emitting regions. The nanoplatelets were further integrated with graphene field-effect transistors, yielding responsivities of $2.5 \times 10^5$ A $W^{-1}$ at 450 nm and $1.9 \times 10^5$ A $W^{-1}$ at 370 nm, respectively. Our method enables scalable growth and selective color tuning of MHP nanoplatelets for on-chip optoelectronic integration.

## 1.1 Introduction

Metal halide perovskite (MHP) nanoplatelets are promising for integrated photonic and optoelectronic applications because of their high photoluminescence quantum yields (PLQY), narrow band emission, tunable band gaps and emission across the visible spectrum, strong light-matter interactions, and long photocarrier lifetimes.[1–5] Their planar geometry, large interfacial area, and strong optical response make them attractive for light-emitting devices, photodetectors, and integration with functional materials. [6,7] Additionally, the emission wavelength of MHP can be precisely tuned by adjusting the halide composition and dimensionality.[4,5,8,9]

Despite these advantages, practical applications of perovskites in optoelectronics are limited by their inherent instability, which arises from their ionic nature and leads to defect formation and ion migration. Defect formation can introduce nonradiative recombination pathways, while ion migration can hinder localized and selective composition tuning.[10,11]Several defect passivation methods have been proposed to suppress nonradiative recombination in MHP, such as surface chemical treatments, core-shell configurations, and ion engineering. Surface chemical treatments are post-synthetic processes in which small molecules or ligands are employed to neutralize trap states, primarily on the surface, thereby improving PL.[12–14] Core-shell structures, on the other hand, are formed during synthesis and involve encapsulating the perovskite core with a protective shell. These heterostructures enhance both environmental and structural stability while also reducing surface defect density. [15–19] Ion engineering enables defect passivation through direct modification of the perovskite lattice, either during synthesis or post-synthesis.[20–25] In-synthesis techniques rely on introducing alkylammonium cations [26–28] and pseudohalides into the MHP lattice.[23,29–31] Incorporation of pseudohalides such as thiocyanate ($SCN^-$),[32,33] formate ($HCOO^-$),[34] acetate ($CH_3COO^-$)[35] and tetrafluoroborate ($BF_4^-$)[36] control the crystallization process, enhance preferential crystal orientation, and assist in defect passivation.[31] Pseudohalides functionalized with carboxyl (O═C–O⁻) groups act as Lewis bases to passivate undercoordinated $Pb^{2+}$ ions, resulting in well-oriented crystals with fewer trap states.[23,37–39] Additionally, fluorine-containing pseudohalides improve the ambient stability and surface hydrophobicity of MHPs.[40] Trifluoroacetate ($TFA^-$) halides, which contain both carboxyl and fluorine functional groups, have emerged as promising candidates to regulate perovskite nucleation, crystallization, and growth orientation.[24,41,42] Previous studies have shown that $TFA^-$ can promote uniform film formation, passivate Pb-related surface and grain-boundary defects, suppress nonradiative recombination, and improve long-term stability.[11,43,44] However, the role of $TFA^-$ as a morphology-directing and surface-passivating agent in promoting well-defined, preferentially oriented perovskite nanoplatelets with improved photophysical properties remains less explored.

While defect control improves the intrinsic optical quality of the nanoplatelets, spatial control over composition and emission is important for integrating different optical responses within a common perovskite platform.[45,46] Post-synthesis ion engineering like anion exchange offers a direct route to tune the optical emission of preformed MHPs because high-quality perovskite structures can first be formed, followed by anion exchange.[8,9] Anion exchange can be performed in the liquid or vapor-phase.[8,47,48] Liquid-phase ion exchange, however, can be limited by precursor solubility, slow exchange kinetics, and degradation of the original MHP morphology. In contrast, vapor-phase exchange avoids direct solvent exposure and therefore offers a promising approach for selective ion exchange and accurate color tuning.[49] Previous vapor-phase studies have mainly focused on graded composition and bandgap control in compact films, or on globally tuning the halide composition of mixed-halide films.[49,50] Nevertheless, selective on-chip color conversion of perovskite, while preserving the original emission of neighboring regions, remains less established and represents an important step toward integrated perovskite platforms with spatially defined optical functionality.

In this study, we combine *in situ* and post-synthesis ion engineering to tune the optical properties and emission of cesium lead bromide ($CsPbBr_3$) nanoplatelets. A single-step solution process enables the direct growth of $CsPbBr_3$@TFA nanoplatelets, using cesium trifluoroacetate (CsTFA). Here, $CsPbBr_3$@TFA denotes TFA-treated nanoplatelets without implying a distinct crystalline shell or substitutional incorporation of TFA into the perovskite lattice. CsTFA serves as both the cesium precursor and pseudohalide source, promoting the formation of well-defined nanoplatelets with preferential crystal orientation while reducing surface defect density. These improvements suppress nonradiative recombination and result in an extended charge-carrier lifetime. We then applied our versatile perovskite-compatible top-down photolithography technique to selectively mask the nanoplatelets,[51,52] followed by spatially controlled vapor-phase anion exchange to produce laterally defined green- and violet-emitting regions on a single chip. These nanoplatelets are subsequently coupled with graphene field-effect transistors (GFETs), showing wavelength-dependent photoresponse. This approach enables scalable *in situ* growth of MHP nanoplatelets with tunable optical properties and supports their integration into optoelectronic platforms.

## 1.2 Results and Discussion

We investigated the influence of $TFA^-$ on the perovskite structure and properties by preparing two types of samples under identical precursor stoichiometry, deposition, and post-annealing conditions. $CsPbBr_3$@TFA nanoplatelets were prepared by spin-coating a precursor solution containing CsTFA and lead bromide ($PbBr_2$). CsTFA facilitates the dissolution of $PbBr_2$ in organic solvents,[11] provides cesium ions ($Cs^+$), and introduces $TFA^-$ as a pseudohalide anion. For comparison, reference $CsPbBr_3$ samples were prepared from a cesium bromide (CsBr) and $PbBr_2$ solution (details are provided in the Experimental Section). A schematic illustration of the growth process for $CsPbBr_3$@TFA nanoplatelets is shown in Figure 1a. The scanning electron microscopy (SEM) images of the $CsPbBr_3$@TFA shown in Figure 1b and Figure S1a (Supporting Information) reveal well-defined rectangular nanoplatelets with an average size of 218 ± 2 nm (Figure S1a-inset, Supporting Information). Atomic force microscopy (AFM) measurements further confirmed the rectangular morphology and a height of 90 nm (Figure S1b,c, Supporting Information). In contrast, $CsPbBr_3$ reference samples exhibited film-like morphology with irregularly shaped grains, as shown in Figure S1 d,e (Supporting Information).

Ultraviolet-visible (UV-Vis) absorption and PL spectroscopy were conducted to investigate the optical properties of the CsTFA/$PbBr_2$ solution, $CsPbBr_3$@TFA nanoplatelets, and $CsPbBr_3$ reference samples. The absorption spectrum of the solution did not exhibit any absorption peaks, indicating the absence of optically detectable preformed perovskite nanoplatelets in solution. This supports that nanoplatelet crystallization occurs during spin-coating and the subsequent annealing process (Figure 1c). $CsPbBr_3$@TFA nanoplatelets showed an intense excitonic peak at 516 nm (Figure 1c), which is slightly blue-shifted to the characteristic absorption edge peak (518 nm) of the $CsPbBr_3$ reference samples (Figure S1f, Supporting Information) and is consistent with the slight bandgap increase observed in the Tauc plot (Figure S2a, Supporting Information). Moreover, Urbach-tail analysis revealed a steeper slope for $CsPbBr_3$@TFA than for $CsPbBr_3$ (Figure S2b, Supporting Information), indicating a sharper optical absorption edge. The extracted Urbach energy decreased from 52.63 meV for the reference $CsPbBr_3$ sample to 27.7 meV for $CsPbBr_3$@TFA. This reduction indicates narrower band-tail broadening and lower energetic disorder in the TFA-derived nanoplatelets, consistent with improved crystallization and reduced defect-related states.[53,54] Furthermore, the nanoplatelets exhibit a narrow PL peak at 520 nm with a full width at half maximum (FWHM) of 14.56 ± 0.5 nm (Figure 1c), slightly blue-shifted from the reference $CsPbBr_3$ emission at 522 nm. The PLQY increased from 0.6% for the $CsPbBr_3$ reference sample to 14.5% for $CsPbBr_3$@TFA ((Figure S2c, Supporting Information), further confirming the improved emissive quality of the TFA-assisted nanoplatelets.

The phase and crystallinity of the nanoplatelets were analyzed using X-ray diffraction (XRD). As shown in Figure 1d, the XRD pattern of $CsPbBr_3$@TFA exhibits intense peaks corresponding to the dominant 3D $CsPbBr_3$ phase (reference code: 00-018-0364),[55] together with additional peaks at 7.33º and 8.14º, suggesting the formation of a TFA-containing low-dimensional perovskite-related phase at the surface.[56–59] Moreover, the secondary phase, $Cs_4PbBr_6$, which typically appears in the XRD pattern of the $CsPbBr_3$ thin films (Figure 1d), was not detected in the nanoplatelets. Angle-dependent grazing-incidence X-ray diffraction (GIXRD) patterns of the (001) plane of the $CsPbBr_3$ phase for both the $CsPbBr_3$ reference and $CsPbBr_3$@TFA nanoplatelets are shown in Figure S3a,b, Supporting Information. The results show that, upon increasing the incident angle (Ψ) from 0.2° to 5°, the diffraction peak of $CsPbBr_3$@TFA nanoplatelets exhibited a smaller shift toward lower 2θ values compared with that of the $CsPbBr_3$ reference sample. In addition, the extracted residual tensile stresses, calculated from the linear relationship between $\sin^2(\Psi)$ and the lattice spacing (*d*) (Figure S3c, Supporting Information),[60,61] were 123 MPa and 54 MPa for the $CsPbBr_3$ reference and $CsPbBr_3$@TFA nanoplatelets, respectively. The smaller peak shift and lower residual tensile stress are consistent with improved preferential orientation and $TFA^-$ coordination at undercoordinated $Pb^{2+}$ sites, which can lower defect-induced lattice distortion in the $CsPbBr_3$@TFA nanoplatelets.[30,62,63]

The local crystal orientations of the $CsPbBr_3$@TFA nanoplatelets and $CsPbBr_3$ reference samples were mapped via electron backscatter diffraction (EBSD). The nanoplatelets aligned preferentially along the [001] crystal direction relative to the sample normal direction, regardless of the crystallinity of the substrate (crystalline Si or amorphous Si/$SiO_2$) (Figure 2). The corresponding pole figures displayed a pronounced central intensity, with maximum multiple of random distribution (MRD) values of approximately 19.50 and 15.99, respectively (Figure 2 b,d). These high MRD values suggest that the nanoplatelets exhibit preferential orientation. In contrast, the EBSD orientation map of the $CsPbBr_3$ reference samples revealed a polycrystalline structure with a variety of grain orientations (Figure S4, Supporting Information). The (001) pole figure exhibited a more uniform distribution, with a lower maximum MRD of approximately 2.12, indicating more random grain orientations. The reliability of the EBSD-derived texture analysis was further validated by generating confidence index (CI) maps with IPF overlays for the same regions (Figure 2 e,f & Figure S4, Supporting Information). These CI maps are displayed on a fixed scale (CI = 0.1-0.3) to enable direct visualization of spatial confidence distribution. Texture calculations were performed only after masking low-confidence and Si-indexed pixels (CI > 0.15, misorientation > 2.0 deg from Si orientation) and using grain-resolved statistics (grain-average orientations). The corresponding CI histograms, representative EBSD patterns, and grain-average misorientation distributions for all datasets used in the analysis are provided in Figures S4 and S5, Supporting Information. These morphological and crystallographic analyses indicated that $TFA^-$ pseudohalides play a key role in regulating the crystal orientation of the perovskite.[42]

The elemental distribution of the nanoplatelets was investigated via high-resolution transmission electron microscopy (HRTEM) and energy-dispersive X-ray spectroscopy (EDXS) mapping. The TEM lamella used for the analysis is shown in Figure S6a (Supporting

Information). The HRTEM image of $CsPbBr_3$@TFA nanoplatelets and the corresponding fast Fourier transform (FFT), taken along the [001] zone axis, are shown in Figure 3a-d. The FFT pattern supports the crystalline nature of the nanoplatelets and indicates the presence of the pure 3D $CsPbBr_3$ phase in the bulk. However, the sensitivity of the nanoplatelets to high-energy electron beams (200 kV) prevented reliable crystalline phase information from being obtained over larger areas (Figure S6b,c, Supporting Information). The cross-sectional HAADF-STEM/EDX maps and line profiles along the arrow (Figure 3e, Figure S6d, Supporting Information) reveal a homogeneous distribution of cesium (Cs), lead (Pb), and bromine (Br) within the bulk of the nanoplatelets. In addition, we observed an intense carbon signal associated with the lamellar coating and a very weak fluorine (F) signal from $TFA^-$. The silicon (Si) and oxygen (O) signals in the EDX map are attributed to the supporting substrate.

The chemical states of C, F, Pb, Br, and Cs at the surface of the $CsPbBr_3$@TFA nanoplatelets and their bonding environments were investigated via X-ray photoelectron spectroscopy (XPS) (Figure 4a-e). All the spectra were corrected using the adventitious C 1s signal at 284.8 eV. The presence of $TFA^-$ on the surface of $CsPbBr_3$@TFA was confirmed through the C 1s and F 1s XPS spectra. The C 1s spectrum showed two distinct peaks at 292.5 eV and 288.8 eV, corresponding to $-CF_3$[64] and −O−C=O groups,[11] respectively (Figure 4a). In the F 1s spectrum in Figure 4b, the peak at 688.7 eV corresponds to $-CF_3$,[11,65]whereas the peak at 683.8 eV corresponds to the F ions bonded to uncoordinated $Pb^{2+}$ on the nanoplatelet surface.[66,67] The Pb 4f spectrum in Figure 4c reveals two dominant peaks for Pb $4f_{7/2}$ and Pb $4f_{5/2}$, which indicate a 0.66 eV shift toward higher binding energies than those of $CsPbBr_3$ thin films. This shift indicates the passivation of uncoordinated $Pb^{2+}$ ions at the nanoplatelet surface.[68,69] Moreover, minor shoulder peaks at ~141 eV and ~137 eV indicate the presence of metallic lead ($Pb^0$).[70] The peak area ratio of $Pb^0$ to Pb4f decreased from 0.12 in the $CsPbBr_3$ reference sample to 0.084 in the $CsPbBr_3$@TFA nanoplatelets, further supporting effective surface passivation by $TFA^-$.[71,72] Similar binding energy shifts of approximately 0.2 eV to 0.4 eV were observed in the Cs 3d and Br 3d spectra (Figure 4d,e). These spectral shifts, combined with the reduction in metallic lead content, suggest that $TFA^-$ may contribute to surface defect passivation[11,68]. Based on the XPS results and TEM EDX data, which did not show significant traces of C or F in the cores of the nanoplatelets, we conclude that $TFA^-$ ions are located on the nanoplatelets surface rather than incorporated into the perovskite crystal structure.[11,44]

We carried out terahertz scattering near-field optical microscopy (THz-s-SNOM) at 0.6 THz to investigate the influence of $TFA^-$ surface defect passivation on the electronic quality of the perovskite nanoplatelets (Figure 5a-c; see details in the Experimental Section).[73] THz-s-SNOM optically scans local THz conductivity via a contact-free method with a nanoscale spatial resolution of 50 nm.[74,75] Experimental near-field images of the $CsPbBr_3$ nanoplatelets on the highly doped silicon substrate (Figure 5a-c) display a negative THz phase $\phi_2$ on the nanoplatelets relative to the highly doped substrate. This is exemplary shown in line profiles for two nanoplatelets (Figure 5d). The response, accompanied by a reduced THz near-field magnitude $S_2$ on the nanoplatelets relative to the substrate, is characteristic of undoped, intrinsic semiconductor materials. The near-field contrasts of the $CsPbBr_3$@TFA nanoplatelets relative to the highly doped silicon substrate were then modeled via the finite-dipole model[76] for the samples [77] (input parameters: tapping amplitude 200 nm, tip radius 40 nm, spheroid length L = 600 nm, film thickness 100 nm) and a Drude conductivity model (effective electron mass = 0.26 $m_0$, where $m_0$ is the electron mass, and the THz permittivity of the lattice is $\varepsilon_L$ = 5.8).[73,78,79] Assuming a carrier mobility of 50 $cm^2V^{-1}s^{-1}$, a typical mobility in lead-halide perovskites,[80] we find that the extracted doping carrier densities are well below $10^{16}$ $cm^{-3}$ (Figure 5e), reflecting the intrinsic semiconducting behavior of the material.

The low degree of doping can be attributed to efficient surface defect passivation, which reduces nonradiative recombination and enhances the emission intensity of $CsPbBr_3$@TFA compared with that of the $CsPbBr_3$ samples (Figure 6a,b, Figure S7 (Supporting Information)). Consistently, the TRPL decay profiles acquired using time-correlated single-photon counting (TCSPC) and fitted using a bi-exponential model (Figure 6c, Table S1, Supporting Note 1) showed longer decay components for $CsPbBr_3$@TFA. The fitted decay components increased from $\tau_1$ = 1.43 ns and $\tau_2$ = 7.59 ns for $CsPbBr_3$ to $\tau_1$= 10.3 ns and $\tau_2$ = 33.24 ns for $CsPbBr_3$@TFA, while the intensity-weighted average lifetime increased from 2.07 ns to 26.38 ns. For the reference $CsPbBr_3$ sample, the late-time region was excluded from the fitting once the signal approached the noise floor. The nonradiative recombination rate constant, estimated from the PLQY and $\tau_{avg}$, decreased from $4.80 \times 10^8$ $s^{-1}$ to $3.24 \times 10^7$ $s^{-1}$ (Table S2, Supporting Note 1), indicating reduced nonradiative losses in $CsPbBr_3$@TFA nanoplatelets.

Beyond improving emission efficiency and suppressing nonradiative recombination, $TFA^-$ surface passivation also enhanced the environmental stability and photostability of $CsPbBr_3$@TFA nanoplatelets. The PL emission remained stable for more than 1440 h under ambient conditions (Figure S8a, Supporting Information), with a gradual increase in PL intensity over time, which can be attributed to the slow passivation effect of $TFA^-$.[81] Photostability was further evaluated under continuous 456 nm laser excitation with an irradiance of $2.83 \times 10^3$ W $cm^{-2}$ under ambient conditions for 8 h (Figure S8b, Supporting Information). The $CsPbBr_3$ reference sample showed pronounced photobleaching, retaining only ~7.5 % of its initial PL. In contrast, $CsPbBr_3$@TFA nanoplatelets retained ~40 % of their initial PL intensity. Building on these results, which support the surface-passivation effect of $TFA^-$ pseudohalides, we propose a surface-passivated structural model for the nanoplatelets, as illustrated in Figure 6d.

Finally, we locally tuned the emission of $CsPbBr_3$@TFA nanoplatelets on-chip by selectively masking them with a Poly(methyl methacrylate)/aluminum oxide (PMMA/$Al_2O_3$) stack using a two-step top-down photolithography process (details in the Experimental Section).[51,52] The process flow is illustrated in Figure 7a. PL spectroscopy was used to monitor the stability of the nanoplatelets during the structuring and encapsulation steps. No changes were observed in the PL spectra (Figure S9, Supporting Information), confirming the stability of the nanoplatelets throughout the patterning process. The chip was subsequently exposed to HCl vapors to initiate a halide exchange process, as schematically shown in Figure 7b (details in the Experimental Section).

The absorption and PL spectra taken at different times during the halide-exchange process show a gradual shift in the absorption edge and emission peak toward shorter wavelengths (Figure S10a, Supporting Information). At short exposure times of 3 and 5 min, two emission peaks are observed, indicating incomplete halide exchange during the early stage of conversion. After 20 min of HCl vapor exposure, the emission peak of $CsPbBr_3$@TFA nanoplatelets was blue-shifted to 413 nm, characteristic of $CsPbCl_3$.[82] The dominant PL peak energy increases from 2.38 eV to 3.00 eV with increasing exposure time, demonstrating exposure-time-controlled spectral tuning during vapor-phase Br-to-Cl exchange (Figure S10b, Supporting Information). XRD analysis of the $CsPbCl_3$@TFA region revealed diffraction peaks consistent with the cubic $CsPbCl_3$ phase (Figure 7c, reference code: 01-084-0437).[83]The diffraction peaks were shifted toward higher 2θ values relative to those to $CsPbBr_3$@TFA, indicating lattice contraction upon substitution of $Br^-$ by the smaller $Cl^-$anion.[84] The XRD results also show that the $CsPbCl_3$@TFA nanoplatelets retain their crystalline phase under ambient conditions for at least 30 days (Figure S11, Supporting Information). Moreover, the Cl 2p XPS spectrum in Figure 7d shows binding energy peaks at 197.5 eV and 199.1 eV, assigned to $Cl2p_{3/2}$ and $Cl2p_{1/2}$, respectively, as expected for $CsPbCl_3$.[85] The depth profile analysis in Figure 7e reveals no detectable residual bromide throughout the etched region, supporting the replacement of bromide with chloride ions.

Next, the PMMA/$Al_2O_3$ stack was removed with toluene. The absorption and PL spectra after resist stripping revealed that the masked regions retained the green emission of $CsPbBr_3$@TFA, whereas the unmasked regions exhibited violet emission from Cl-rich $CsPbCl_3$@TFA (Figure 7f). The optical micrograph in Figure 8a shows representative patterned nanoplatelet features on the chip, while the confocal PL maps demonstrate their spatially resolved emission (Figure 8b-f). The masked and unmasked regions showed well-defined green-and violet-emitting features, respectively. PL line-profile analysis further confirmed the optical uniformity and spatial confinement of both emission regions (Figure S12a,b, Supporting Information). The corresponding emission-intensity profiles confirmed the uniformity of the converted emission across the unmasked and masked areas (Figure S12c,d, Supporting Information). Additional confocal mapping at a representative patterned feature was used to assess lateral halide diffusion (Figure S12e,f, Supporting Information). The wavelength-resolved confocal images in Figure S11f show that the violet-emitting exchanged regions and the green-emitting regions remain spatially separated across the patterned interface. Confocal PL maps and spectra collected at the boundary between the unmasked and masked regions further revealed the coexistence of green- and violet-emitting domains on the chip (Figure 8g, h). The absence of noticeable color mixing across the boundary indicates limited lateral diffusion under the present exchange conditions. These results demonstrate the compatibility of top-down patterning with selective and localized color tuning on a single chip.

We next integrated the $CsPbBr_3$@TFA and $CsPbCl_3$@TFA nanoplatelets with GFETs to realize hybrid phototransistors (see Experimental Section for details). Raman spectra of bare GFET,

$CsPbB_3$@TFA-graphene, and $CsPbCl_3$@TFA-graphene were collected at different locations across the samples (Figure S13, Supporting Information). The characteristic graphene G band at ≈1590 $cm^{-1}$ and 2D band at ≈2680 $cm^{-1}$ are retained after both perovskite integration and halide exchange. In addition, no noticeable increase in the D band is observed after deposition of $CsPbBr_3$@TFA and after conversion to $CsPbCl_3$@TFA, indicating that these processing steps do not introduce significant additional disorder into the graphene lattice. The influence of the perovskite layer on graphene transport was then examined using dark transfer characteristics. As shown in Figure S14, pristine graphene exhibits a clear ambipolar response with a CNP (charge-neutrality point) at around 40 V. After integration with $CsPbBr_3$@TFA, the CNP shifts toward lower gate voltage, accompanied by an increase in drain current. This negative CNP shift indicates n-type doping of the graphene channel, consistent with electron transfer from the $CsPbBr_3$@TFA layer to graphene after spin coating.[86] After conversion to $CsPbCl_3$@TFA, the device still preserves ambipolar transport, but with a CNP shift towards positive gate voltages, indicating p-type doping of graphene. These opposite CNP shifts suggest that the perovskite composition modifies the electronic environment of graphene while maintaining gate-controlled transport through the channel. The photoresponse of the hybrid phototransistors was further evaluated from transfer curves measured under 520 nm, 450 nm, and 370 nm illumination with a fixed light intensity of 8.39 W $m^{-2}$. As shown in Figure 9a, both $CsPbBr_3$@TFA-graphene and $CsPbCl_3$@TFA-graphene devices exhibited light-induced modulation of the transfer curves compared with the dark state. The positive shift of the Dirac point under illumination indicates hole transfer from the perovskite to graphene, while electrons remain trapped in the perovskite. These trapped electrons act as an additional local gate, giving rise to a photogating effect that modulates the graphene carrier density and drain current. [86] This behavior can be explained by a photogating effect at the graphene-perovskite interface,[87,88] where photogenerated carriers are separated under illumination.

The wavelength-dependent transfer curves also reveal composition-controlled spectral tuning after halide exchange. The $CsPbBr_3$@TFA-graphene device showed stronger modulation under visible illumination at 520 nm and 450 nm, which is consistent with the visible-light absorption of $CsPbBr_3$@TFA, whereas the $CsPbCl_3$@TFA-graphene device responded more strongly at 370 nm, consistent with the wider bandgap of $CsPbCl_3$@TFA. The corresponding current-voltage and photocurrent characteristics (Figure 9b and Figure S15, Supporting Information), suggest efficient collection of photogenerated carriers through the graphene channel. Moreover, Responsivity and external quantum efficiency (EQE) were extracted at incident power densities of 1.69 W $m^{-2}$, 4.20 W $m^{-2}$, and 8.39 W $m^{-2}$, at $V_g$ = 40 V and $V_{ds}$ = 1 V. Both devices showed wavelength- and power-dependent responsivity, with higher values at lower illumination intensity, consistent with photogating behavior in graphene-semiconductor hybrid phototransistors.[87] The $CsPbBr_3$@TFA-graphene and $CsPbCl_3$@TFA-graphene devices reached maximum responsivities of 2.5 $\times10^5$ A $W^{-1}$ and 1.9 $\times10^5$ A $W^{-1}$, respectively, at 1.69 W $m^{-2}$ (Figure 9c), comparable to reported graphene-perovskite phototransistors.[86,89] A similar trend was observed for EQE (Figure 9d). These results demonstrate that selective and localized on-chip halide conversion enables wavelength-selective phototransistor operation while preserving gate-controlled graphene transport.

**Conclusion**

In summary, this study demonstrated a combined *in situ* and post-synthesis ion-engineering to enable the growth of $CsPbBr_3$@TFAnanoplatelets and their selective color conversion on-chip. We used CsTFA as a source of $TFA^-$ pseudohalides, and the resulting nanoplatelets exhibited emission at 520 nm with a narrow linewidth 14.56 ± 0.5 nm. XPS analysis revealed that $TFA^-$ facilitated the passivation of uncoordinated $Pb^{2+}$ and reduced the metallic lead content, thereby suppressing nonradiative recombination, enhancing the PLQY, and prolonging the average PL lifetime to 26.38 ns. Furthermore, structural analyses indicated $TFA^-$ reduced lattice distortion and promoted the preferential orientation of the nanoplatelets along the [001] crystal direction, regardless of the substrate crystallinity. A spatially selective post-synthesis color conversion approach was then employed to shift the nanoplatelet emission from green to violet. A two-step photolithographic process was used to selectively cover predefined regions of the on-chip nanoplatelets with a PMMA/$Al_2O_3$ mask. Subsequent exposure to HCl vapor locally shifted the PL emission from 520 nm to 413 nm in the unmasked regions, producing laterally defined $CsPbBr_3$@TFA and $CsPbCl_3$@TFA regions on the same substrate. The $CsPbBr_3$@TFA and $CsPbCl_3$@TFA nanoplatelets were integrated with GFETs and showed wavelength-dependent photoresponse, with responsivities of $2.5 \times 10^5$ A $W^{-1}$ at 450 nm and $1.9 \times 10^5$ A $W^{-1}$ at 370 nm. This work provides a practical route for integrating defect-passivated, composition-tunable perovskite nanoplatelets into wavelength-selective on-chip optoelectronic devices.

**Experimental**

**<u>$CsPbBr_3$@TFA nanoplatelets and $CsPbBr_3$ thin film deposition:</u>** Si and Si/$SiO_2$ substrates were sequentially sonicated in acetone and isopropyl alcohol for 10 minutes each. Afterward, they were blow-dried with nitrogen and treated with oxygen plasma (Tepla, Semi 300) for 10 minutes. The cleaned substrates were then transferred to a nitrogen-filled glove box. For $CsPbBr_3$@TFA nanoplatelet deposition, cesium trifluoroacetate (CsTFA, Thermo Scientific) and lead bromide ($PbBr_2$, ≥98%, Sigma-Aldrich) were dissolved in anhydrous DMSO at a molar ratio of 1.1:1. For reference $CsPbBr_3$ thin-film deposition, cesium bromide (CsBr, >99.0%, TCI) and $PbBr_2$ were dissolved in anhydrous DMSO at the same CsBr to $PbBr_2$ molar ratio of 1.1:1. Both precursor solutions were prepared with total solids concentrations of approximately 12 wt%. Both precursor mixtures were stirred overnight at 60 °C and then filtered through 0.2 µm polytetrafluoroethylene (PTFE) filters before deposition. Before deposition, we maintained both the solution and the substrates at 60 °C. We spin-coated 80 µL of the corresponding precursor solution onto each 4 $cm^2$ substrate at 2000 rpm for 40 seconds, followed by annealing at 80 °C for 10 minutes. The glovebox $O_2$/$H_2O$ level during the deposition was 3.2 ppm /2.7 ppm.

**<u>Top-down patterning of $CsPbBr_3$@TFA nanoplatelets on chip:</u>**

A perovskite-compatible top-down patterning process utilizing a double-stack resist consisting of a poly(methyl methacrylate) (PMMA) protective layer and an AZ Mir 701 positive-tone photoresist was conducted as described in previous work.[51,52] The samples were exposed to ultraviolet (UV) light for 25 s via a contact lithography mask aligner (EVG 420) system. The samples were then baked for 90 s at 115°C. The patterns were developed in an MF26A developer for 35 s, followed by rinsing with deionized water for 7 seconds. The samples were subsequently etched via plasma with $BCl_3$ and HBr gases via a PlasmaLab System 100 inductively coupled plasma (ICP)-RIE tool (Oxford Instruments). Finally, the resist stack was removed by immersing the samples in toluene at 80°C for 1 hour.

**Selective halide exchange:**

Photolithography and etching steps were performed to define masked and unmasked regions. A PMMA (350–400 nm)/evaporated $Al_2O_3$ (5 nm)/AZ Mir 701 positive-tone photoresist stack served as a mask for certain perovskite arrays during the anion exchange process. Anion exchange was performed by exposing the unmasked $CsPbBr_3$@TFA nanoplatelets to chlorine vapors, which was achieved by heating hydrochloric acid (HCl) at 108°C in a sealed vial for 20 minutes.

**Fabrication of perovskite-graphene phototransistors:** GFETs were fabricated on 2 × 2 $cm^2$ Si/90 nm $SiO_2$ substrates using conventional photolithography. Commercially available, large-area chemical-vapor-deposited graphene on copper was transferred onto the substrates using a PMMA-assisted transfer method. The graphene was patterned by oxygen plasma-based RIE, and top contacts were fabricated by sputtering 50 nm Ni, followed by a lift-off process. To fabricate a hybrid phototransistor, $CsPbBr_3$@TFA nanoplatelets were spin-coated onto the GFET, followed by the vapor-phase halide exchange process. All measurements were performed under ambient conditions at room temperature.

**Characterization***:* Scanning electron microscopy (SEM) measurements were performed in a Zeiss SUPRA 60 at 4 kV with a working distance of 4 mm. Atomic force microscopy (AFM) images were taken using a Bruker Dimension Icon instrument in tapping mode. Electron backscatter diffraction (EBSD) measurements were performed on an FEI Verios 460L SEM with an EDAX Clarity direct electron detection system. The sample was held at a 70° tilt angle. An accelerating voltage of 8 kV was used with a beam current of 400 pA. Electron diffraction patterns were collected with a step size of 25 nm and a detector exposure time of 15 ms. The resulting diffraction patterns were post-processed and re-indexed using spherical indexing (bandwidth of 255) in OIM analysis 9. For spherical indexing, a simulated master pattern of cubic $CsPbBr_3$ (space group *Pm-3m*) was used. For pole figure plotting, grain CI standardization was performed; only grain average orientations corresponding to $CsPbBr_3$ were plotted. To ensure that EBSD texture metrics reflect reliably indexed data, low-confidence and Si-indexed pixels were excluded using a fixed confidence-index threshold (CI > 0.15, misorientation > 2.0 deg from Si orientation) before texture evaluation. Texture was computed using grain-resolved statistics (grain-average orientations) to reduce sensitivity to isolated pixels and local indexing outliers. CI maps in the main figure are displayed over a fixed range (0.1 - 0.3) for visualization and comparison of spatial confidence distribution. Indexing robustness is documented via CI histograms and grain-average misorientation distributions. To accommodate low dose EBSD imaging, a highly sensitive direct electron detector was used (EDAX clarity). Experiments were performed on conductive Si substrates or $Si/SiO_2$ (90 nm) with copper clamps contacting the perovskite surface. No degradation or charging of the samples was observed.

Ultraviolet-visible (UV-Vis) spectroscopy was performed using a PerkinElmer UV-Vis spectrophotometer. Photoluminescence (PL) mapping was conducted via a WiTec confocal Raman microscope. The $CsPbBr_3$ samples were scanned with a 457 nm CW laser at 1 µW power. For the $CsPbCl_3$ samples, PL measurements were performed with a CW, diode-pumped, frequency-tripled solid-state laser ($\lambda$ = 355 nm). The emitted light was coupled into a monochromator (Princeton Instruments, Acton SP2500, gratings: 300 lines mm$^{-1}$), and the spectrally dispersed light was detected by a thermoelectrically cooled charge-coupled device camera (Princeton Instruments). The measurements were performed at room temperature under ambient conditions. XRD spectra with filtered Cu-Kα radiation (wavelength of 1.5405 Å) were taken via a PANalytical instrument at a current of 40 mA and a voltage of 40 kV. Grazing Incidence XRD (GIXRD) was performed at various tilt angles. The TEM sample was prepared with a focused ion beam (FIB) Strata FIB 400. HRTEM and EDXS elemental mappings were performed with a JEOL JEM F200 at 200 kV. X-ray photoelectron spectroscopy (XPS) measurements were performed using a Thermo Scientific KA1066 system equipped with monochromatic Al K-α radiation (1486.6 eV). The spectra were charge-corrected by referencing the C–C peak of adventitious carbon to 284.8 eV to account for possible surface charging. Signals from C1s, O1s, F1s, Si2p, Br3d, Cl2p, Cs3d, and Pb4f were recorded, and elemental ratios were quantified from the integrated peak areas using appropriate sensitivity factors. Depth profiling was carried out via $Ar^+$ ion sputtering. THz measurements were performed via custom-built all-electronic THz-sSNOM. 600 GHz radiation from a Schottky diode-based electronic multiplier chain is scattered at the tip of a conductive atomic force microscopy (AFM) cantilever (RMN 25Pt200B−H, 40 nm tip radius) and detected in the far field by a corresponding heterodyne detector. Absolute PLQY measurements were performed using a Hamamatsu C9920-03 absolute PL quantum-yield measurement system. As an excitation source, a 150 W xenon lamp (Hamamatsu L8474), mounted in a Hamamatsu E7536 lamp housing equipped with a starter and operated using a Hamamatsu xenon/mercury-xenon lamp power supply, was used. The broadband xenon-lamp emission was passed through a manually controlled Hamamatsu A10080-01 monochromator to obtain excitation light centered at 350 nm. The selected excitation light was transferred through a fiber probe and an A10093 excitation adapter to the sample positioned inside a 3.3-inch Spectralon-coated Hamamatsu A10094 integrating sphere. The integrating sphere collected the photoluminescence emitted by the sample together with the unabsorbed and scattered excitation light. The collected optical signal was transferred through a Hamamatsu A10079-01 light guide to a PMA-12 Hamamatsu C10027 photonic multichannel analyzer equipped with a cooled 1024-channel CCD detector. The recorded spectra were analyzed using Hamamatsu U6039-05 quantum-yield measurement software, and the absolute PLQY was determined from the ratio of the spectrally integrated number of emitted photons to the number of photons absorbed by the sample.

Time-correlated single-photon counting (TCSPC) measurements were performed using a custom-built setup based on an inverted microscope (TE300, Nikon), following our previously reported configuration.[52] A confocal scan head (PMC2000, Nikon) was used for scanning and collection. The samples were excited using a pulsed blue laser diode module (PicoQuant) with a wavelength of 450 nm, pulse duration below 200 ps, repetition rate of 10 MHz, and average power of 1 mW. The emitted PL was collected through the same confocal scan head and detected using a single-photon-sensitive photomultiplier system (PMC 100, Becker & Hickl). An optical bandpass filter (500-540 nm) was used in the detection path to suppress the excitation signal. The temporal resolution was better than 1 ns, and the instrument response function (IRF), measured using scattered laser light, had a full width at half maximum of 600 ps.

Confocal spectral fluorescence imaging was performed using an LSM 980 confocal microscope with a 34-channel Quasar detector (Zeiss, Oberkochen). The samples were excited with a 405 nm diode laser, and the emitted light was collected with an EC Plan-Apochromat 10x/0.45 objective. Images were acquired in l-mode. The microscope system was controlled with the ZEN blue software, version 3.13.109.05000 (64-bit). The phototransistors were characterized in a Cascade Summit 12000 Probe Station with a Hewlett Packard Precision semiconductor parameter analyzer A4156A under dark conditions and illumination from UV (370 nm), blue (450 nm), and green (520 nm), light-emitting diodes (LEDs) at different incident power densities. The responsivity (R (A $W^{-1}$ )) was calculated using the formula R= $I_{ph}/(I_{illumination} \times A)$, where $I_{photo}$ (A), $I_{illumination}$ (W $m^{-2}$), and A ($9.6\times10^{-10}$ $m^2$) represent photocurrent, illumination power density, and illuminated photoactive area, respectively.[90,91] The external quantum efficiency (EQE) was extracted using the relation EQE = R × (h × c/q × λ), where h is Planck's constant, c is the speed of light, q is the elementary charge, and $\lambda$ is the wavelength.[92]

**Acknowledgments:** This project has received funding from the German Research Foundation (DFG) through the project Hiper-Lase (GI 1145/4-1, LE 2440/12-1, HA 3022/13-1, and RI1551/18-1), the European Union's Horizon 2020 research and innovation programme under the project FOXES (951774), the Deutsche Forschungsgemeinschaft within TRR 404 Active-3D (project number: 528378584), and Federal Ministry of Research, Technology and Space (BMFTR) through the project NEPOMUQ (13N1711A and 01N124211B). The authors thank P. Grewe and Dr. U. Böttger from the Electronic Material Research Lab, RWTH Aachen University, for their support with XRD measurements and analysis. The authors also thank Dr. Sabrina Ernst from the Light Microscopy Facility, a Core Facility of the Interdisciplinary Center for Clinical Research (IZKF) Aachen within the Faculty of Medicine at RWTH Aachen University, for her support with confocal microscopy.  L.S. and B.E. acknowledge the Dutch Research Council (NWO), Gatan (EDAX), Amsterdam Scientific Instruments (ASI) and CL Solutions for financing the project 'Achieving Semiconductor Stability From The Ground Up (NWO project number 19459).

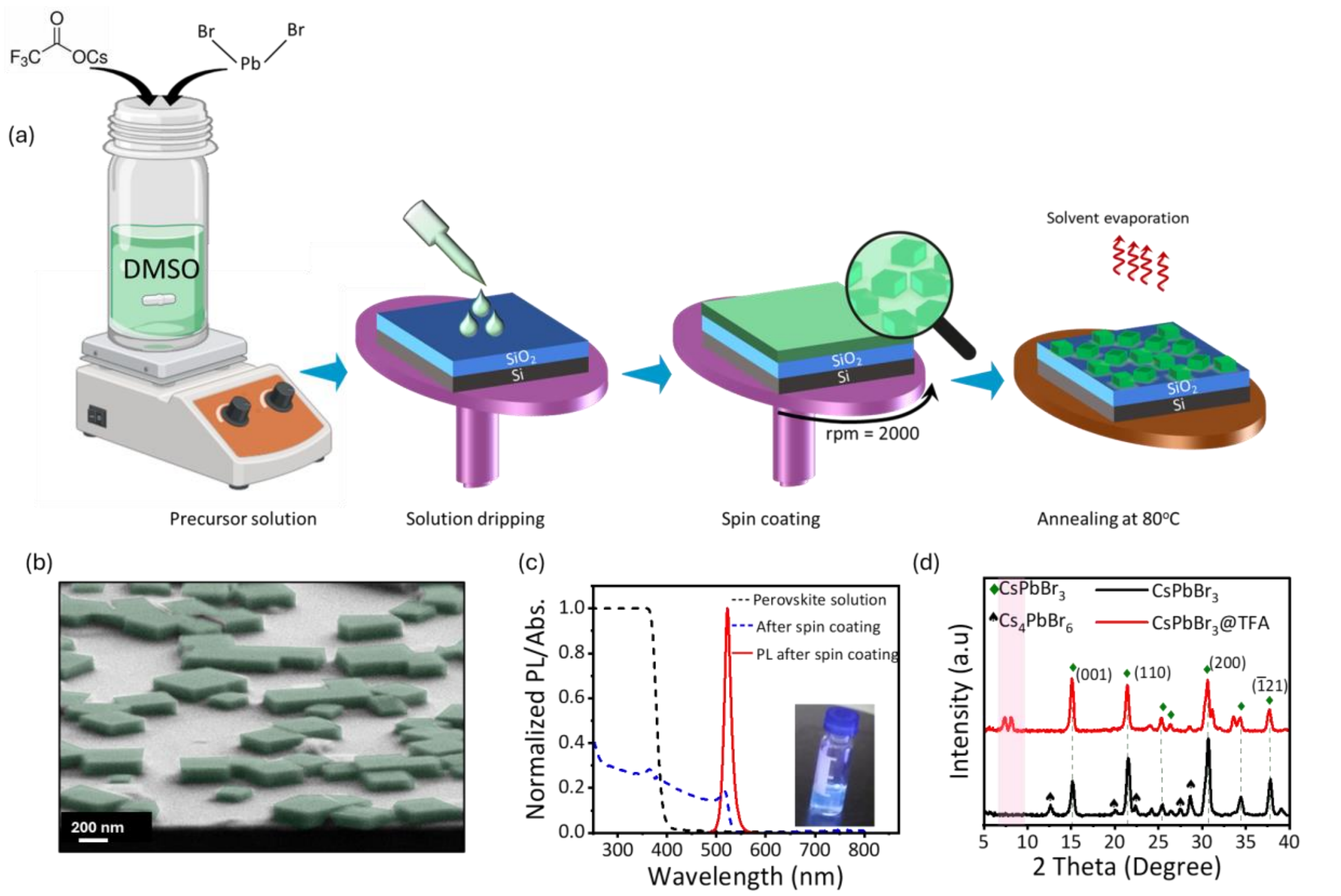

**Figure 1.** Synthesis of $CsPbBr_3$@TFA nanoplatelets. (a) Schematic illustration of the *in situ* fabrication of $CsPbBr_3$@TFA nanoplatelets via a one-step spin-coating process. (b) False-color tilted top-view SEM image of the nanoplatelets. (c) Optical characterization of the nanoplatelets; the dashed black line shows the absorption spectrum for the CsTFA/$PbBr_2$ solution, the dashed blue line shows the absorption spectrum of the as-grown nanoplatelets, and the solid red line shows the PL peak position of the nanoplatelets. (e) XRD patterns of the $CsPbBr_3$@TFA nanoplatelets and $CsPbBr_3$ reference samples.

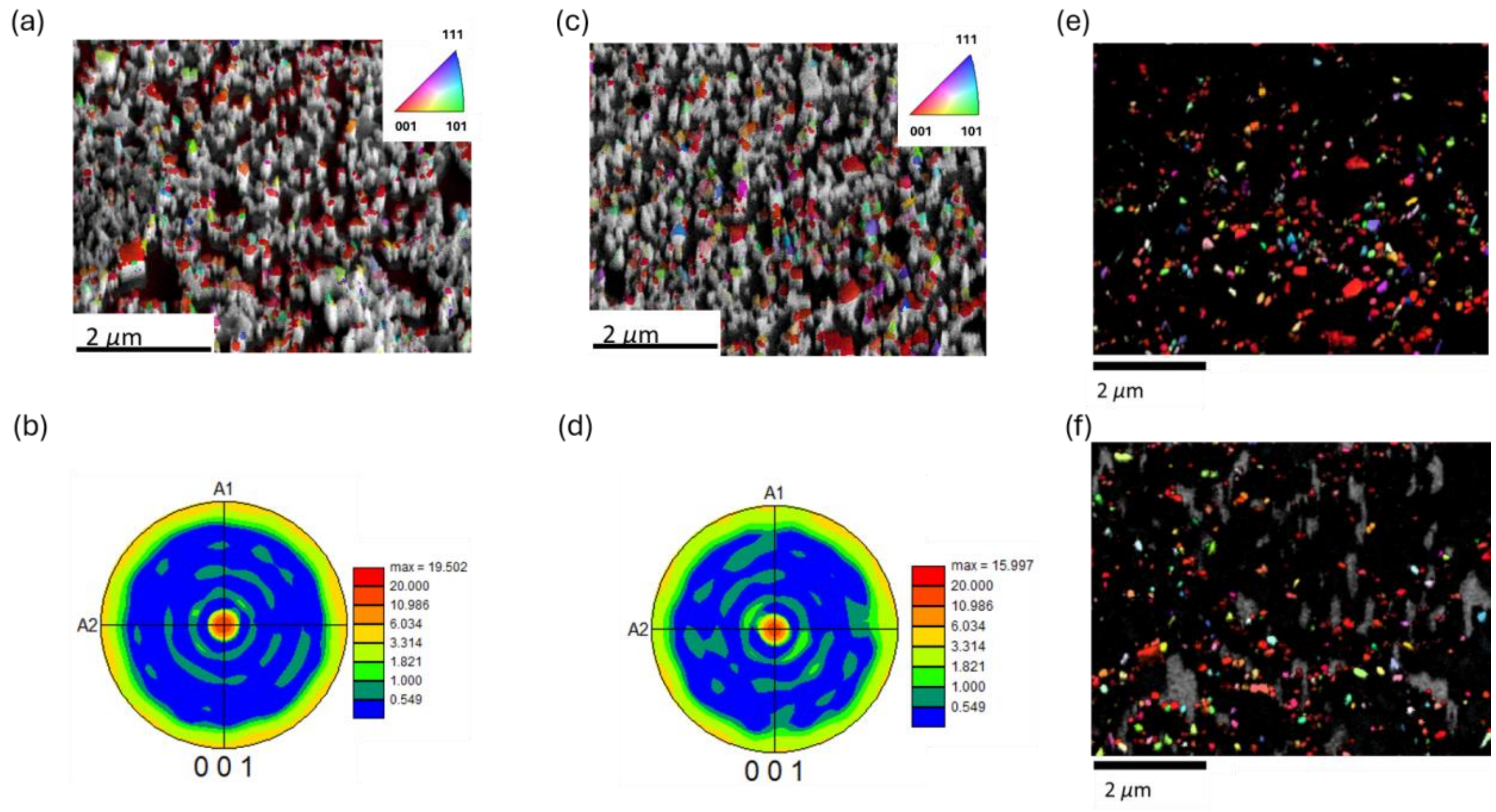

**Figure 2**. EBSD analysis of nanoplatelets on different substrates. Inverse pole figure (IPF) maps (a, c) and corresponding (001) pole figures (b, d) for nanoplatelets deposited on (a, b) crystalline Si and (c, d) amorphous Si/$SiO_2$ substrates. The IPF maps reveal that the nanoplatelets exhibit a strong preferential alignment along the [001] crystal direction regardless of substrate crystallinity. (e, f) confidence-index (CI) maps with IPF overlay for the same regions, illustrating the spatial distribution of indexing confidence in the data used for texture evaluation.

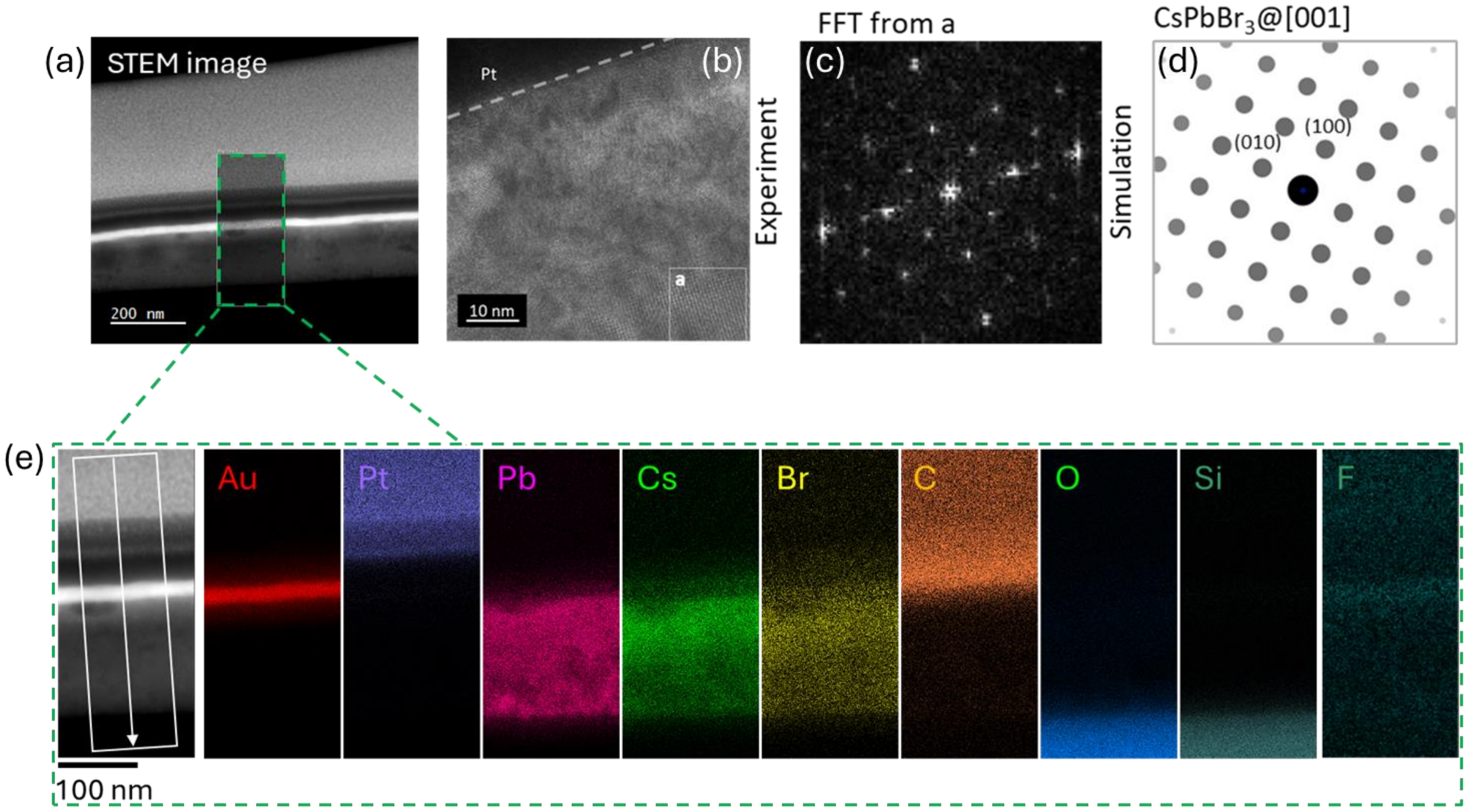

**Figure 3.** TEM characterization of nanoplatelets. (a-c) HRTEM images and FFT pattern of nanoplatelets along the [001] zone axis line. (d) Simulated diffraction pattern. (e) TEM-EDX mapping of the nanoplatelets.

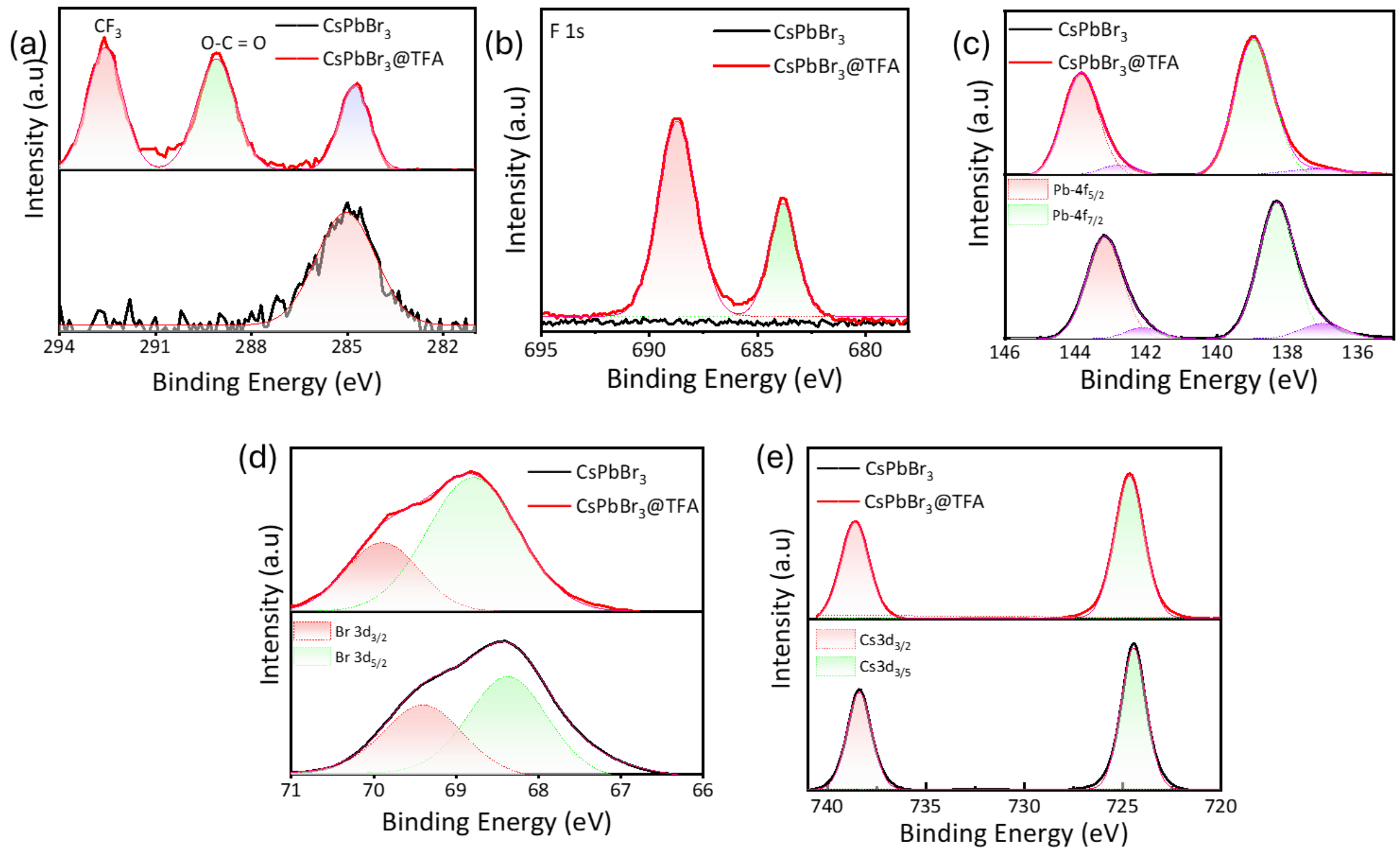

**Figure 4.** High-resolution XPS spectra of key elements in the $CsPbBr_3$ reference sample and $CsPbBr_3$@TFA nanoplatelets. (a) C 1s spectrum, showing the presence of surface-bound organic species; (b) F 1s spectrum, confirming the successful incorporation of TFA; (c) Pb 4f, (d), Br 3d, and (f) Cs 3d spectrum, revealing shifts in binding energies indicative of altered chemical environments following TFA surface modification.

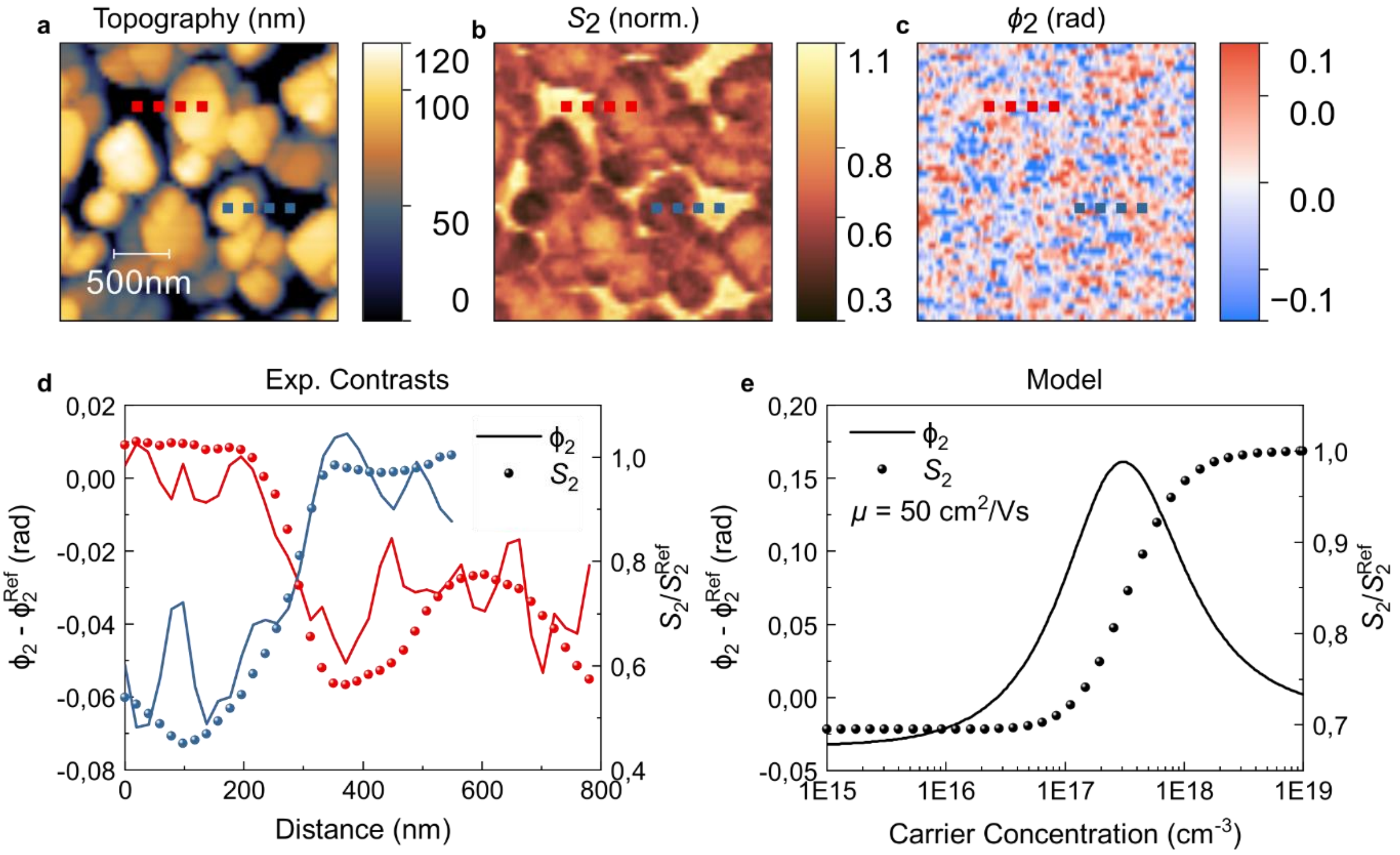

**Figure 5.** THz nanoimaging of $CsPbBr_3$@TFA nanoplatelets. (a-c) THz nanoimaging of $CsPbBr_3$@TFA nanoplatelets on a highly doped silicon substrate. (d) Normalized phase $\phi_2$ (lines) and magnitude $S_2$ (dots) near-field profiles along two different nanoplatelets and the highly doped substrate indicated by the red and blue dashed lines in a-c. (e) Modeled near-field contrasts $\phi_2$ (lines) and $S_2$ (dots) at 600 GHz assuming an intrinsic charge carrier mobility of 50 $cm^2$/Vs.

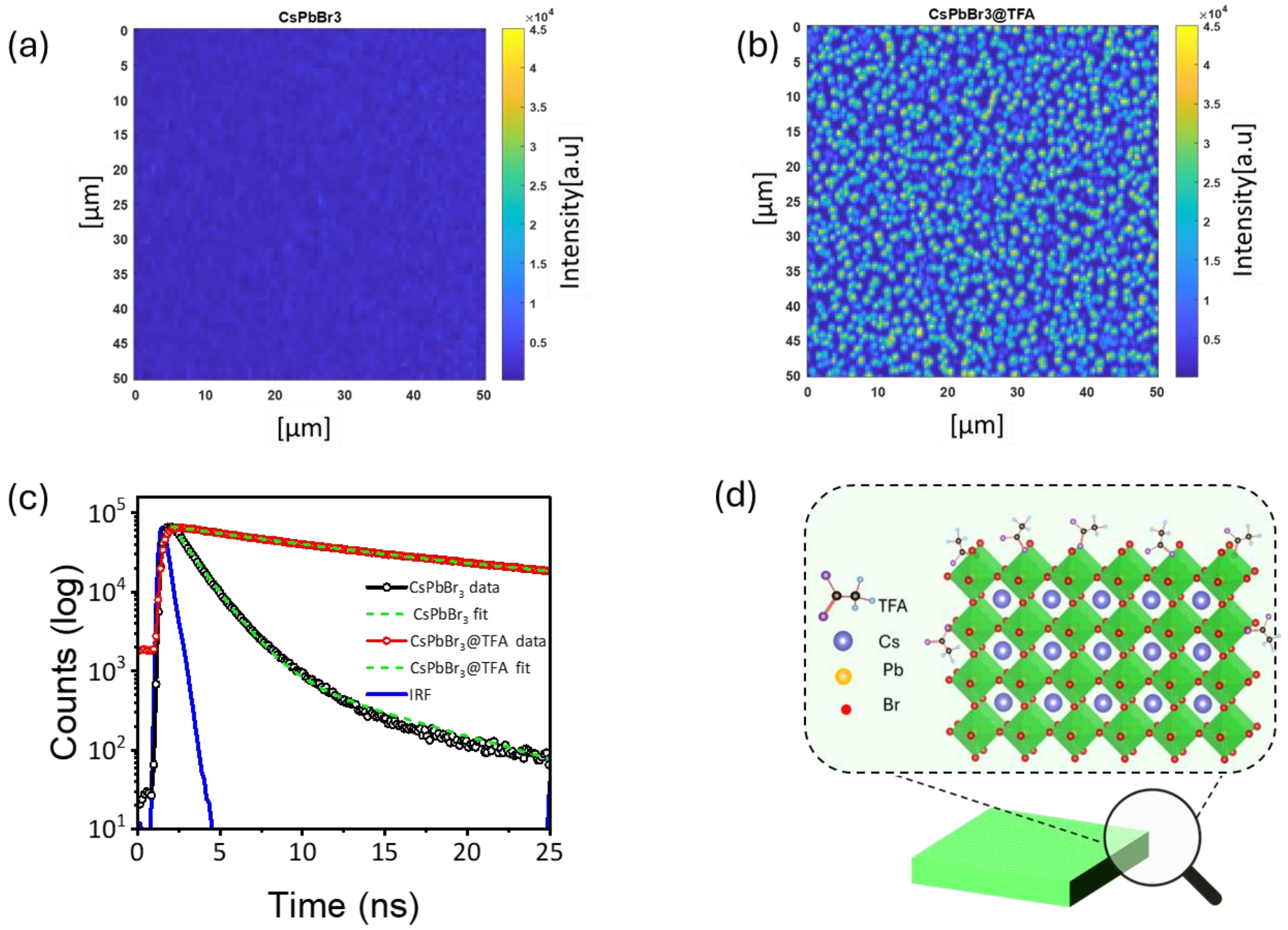

**Figure 6.** PL characterization of perovskite samples. PL area scans for the comparison of average PL intensity for (a) $CsPbBr_3$@TFA nanoplatelets and (b) $CsPbBr_3$ thin films. (c) TRPL decay profiles of $CsPbBr_3$ and $CsPbBr_3$@TFA, shown with the corresponding bi-exponential fits and the instrument response function (IRF). The late-time region of reference $CsPbBr_3$ was not analyzed further once the signal approached the noise floor. (d) Schematic illustration of the proposed surface-passivated $CsPbBr_3$@TFA nanoplatelets.

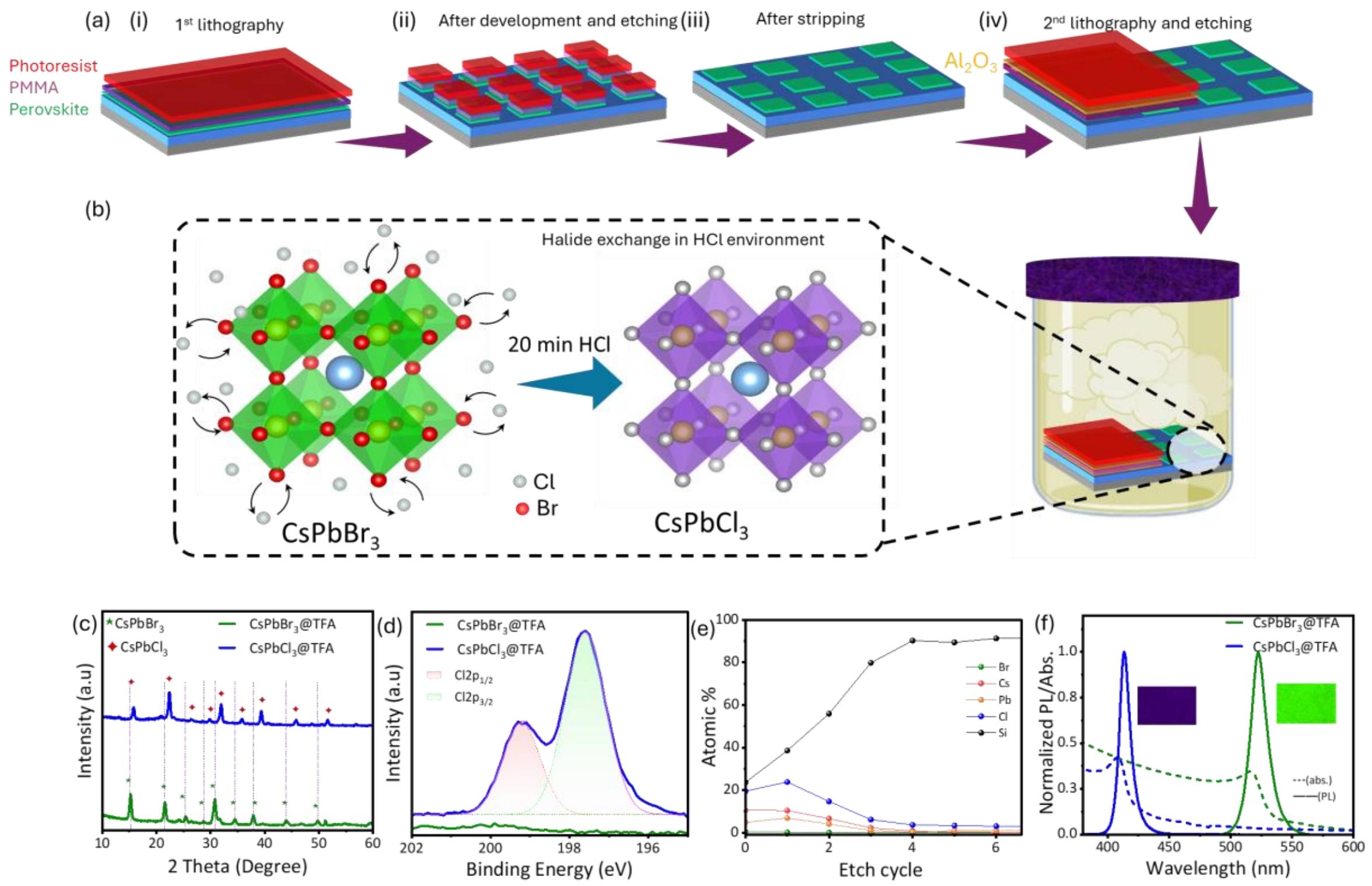

**Figure 7**. Fabrication and halide-exchange characterization of $CsPbBr_3$@TFA nanoplatelets (a) Schematic illustration of the critical fabrication steps, including top-down photolithography. (b) Illustration of nanoplate treatment with HCl vapors and the possible halide exchange mechanism. (c) XRD patterns of the samples before and after halide exchange. (d) XPS peak position for Cl 2p before and after halide exchange. (e) Depth profile extracted from XPS data. (f) PL (solid lines) and UV-Vis absorption (dashed lines) of samples treated for different durations in a chlorine environment

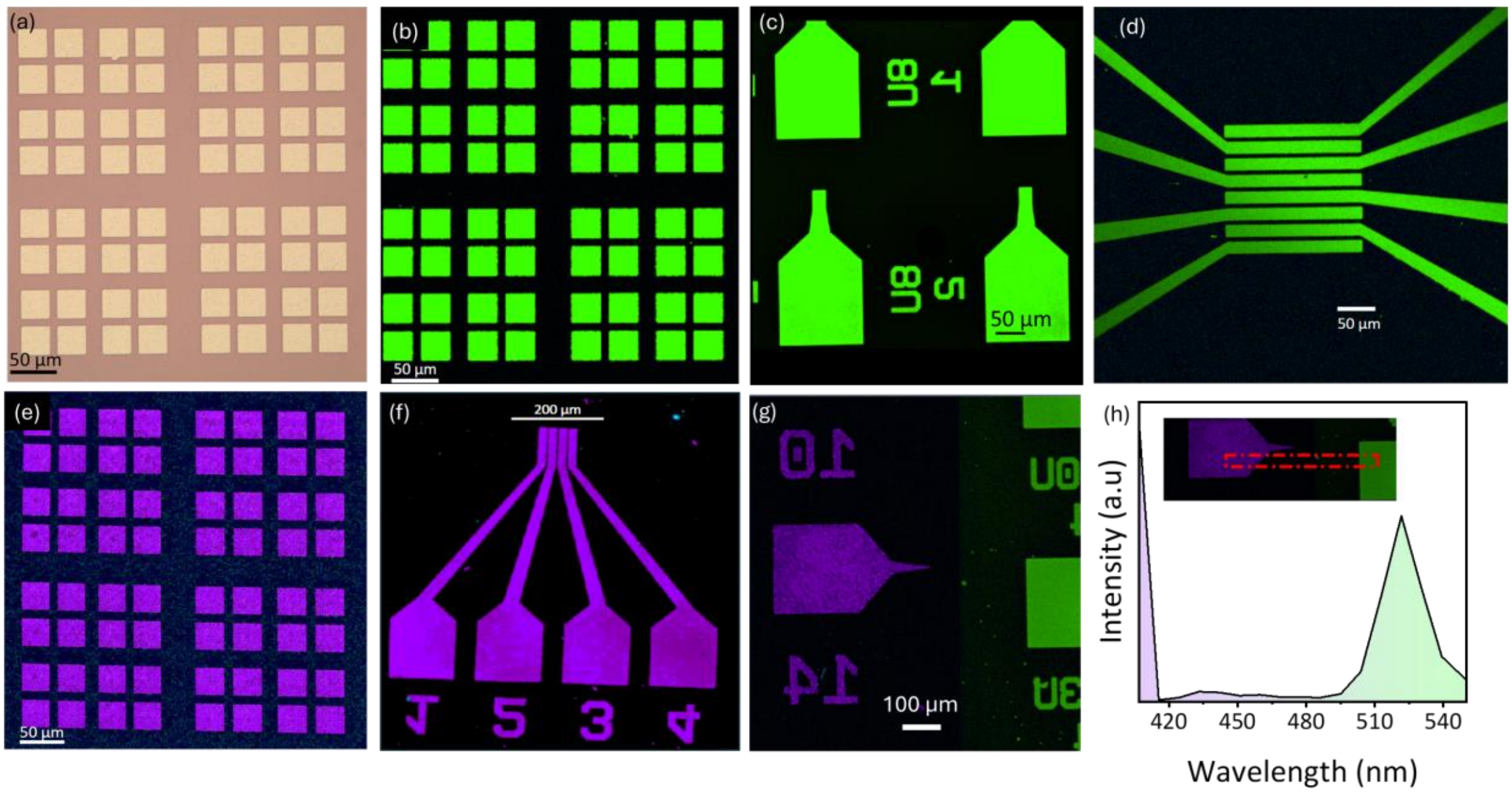

**Figure 8.** Optical and confocal imaging of patterned nanoplatelets. (a) Optical micrographs of representative features achieved after top-down photolithography. (b-d) Confocal PL images of patterned areas in the masked regions of the chip showing retained green emission. (e,f) Confocal PL images of patterned areas in the unmasked regions after HCl vapor exposure, showing violet emission. (g,h) Confocal PL maps and corresponding emission spectra acquired from a single chip after halide exchange.

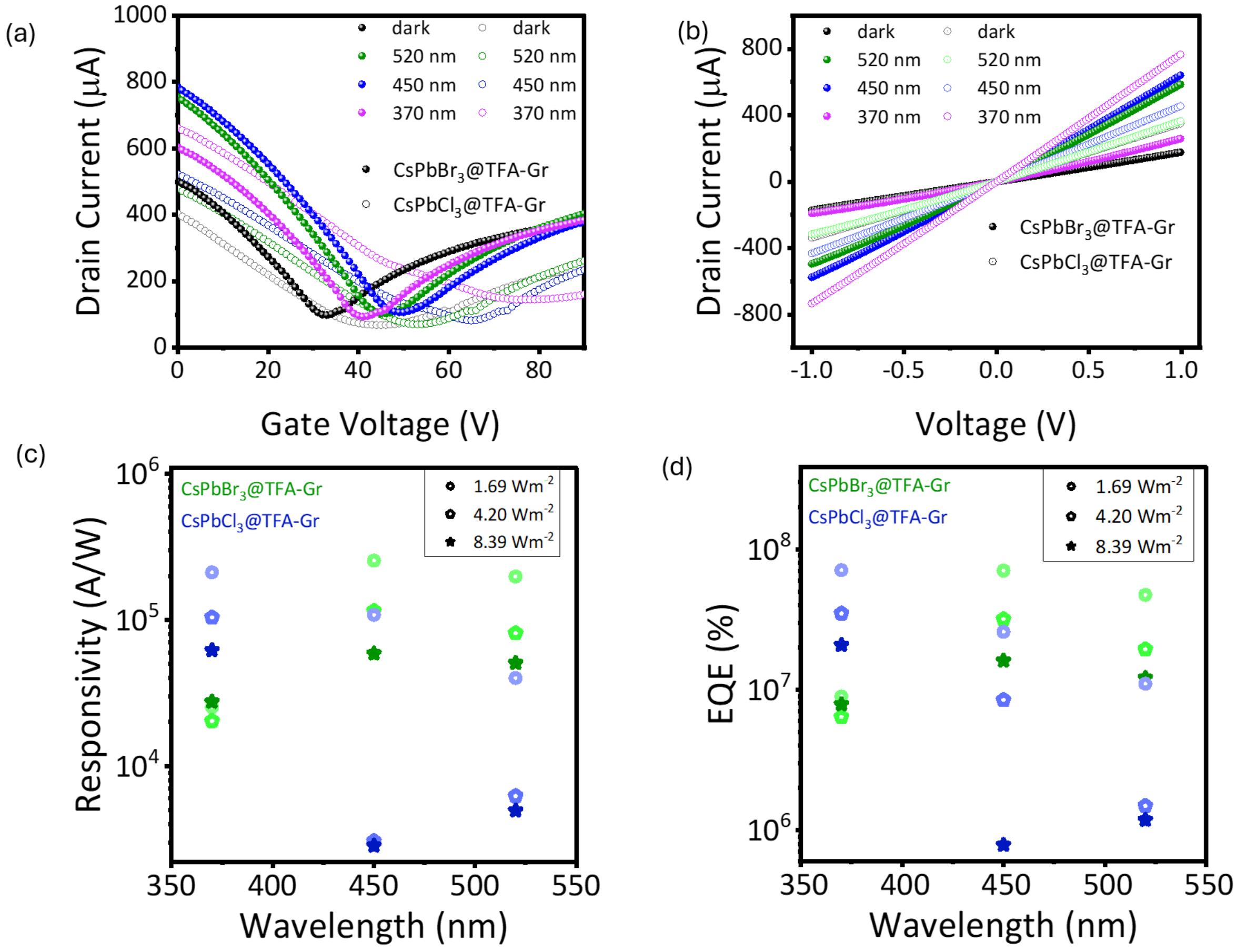

**Figure 9.** Optoelectronic performance of hybrid phototransistors. (a) Transfer characteristics of the hybrid FETs, measured under dark and illuminated conditions at excitation wavelengths of 520 nm, 450 nm, and 370 nm. The illumination intensity is kept at 8.39 W $m^{-2}$ (b) Output characteristics of the hybrid phototransistor at a fixed $V_g$ of 40 V under excitation wavelengths of 520 nm, 450 nm, and 370 nm. (c,d) Wavelength- and intensity-dependent responsivity and EQE of $CsPbBr_3$@TFA-graphene and $CsPbCl_3$@TFA-graphene phototransistors measured at incident light intensities of 1.69 W $m^{-2}$, 4.20 W $m^{-2}$, and 8.39 W $m^{-2}$. The measurements were acquired at $V_g$ = 40 V and $V_{ds}$ = 1 V.

**Supporting Information**

**Facile On-Chip Growth and Spatially Selective Color Tuning of $CsPbBr_3$@TFA Nanoplatelets via Ion Engineering**

Sana Khan[1,2], Saeed Goudarzi[2], Lars Sonneveld[3], Felix Thiel[4], Stephan Schäffer[4], Bart Macco[5], Ke Ran[1,6,7], Sercan Ozen[8], Naho Kurahashi[9,10], Peter Haring Bolívar[4], Felix Lang[8], Bruno Ehrler[3], Burkay Uzlu[1], Thomas Riedl[9,10], Gerhard Müller-Newen[11], Surendra. B. Anantharaman[1,12], Maryam Mohammadi[1,*], Max. C. Lemme[1,2,*]

[1]*AMO GmbH, Otto-Blumenthal-Straße 25, 52074 Aachen, Germany*

[2]*RWTH Aachen University, Chair of Electronic Devices, Otto-Blumenthal-Str. 25, 52074 Aachen, Germany*

[3]*LMPV-Sustainable Energy Materials Department, AMOLF Science Park 104, 1098 XG Amsterdam, The Netherlands*

[4]*Institute for High Frequency and Quantum Electronics, University of Siegen, 57076 Siegen, Germany*

[5]*Department of Applied Physics and Science Education, Eindhoven University of Technology, P.O. Box 513, 5600, MB, Eindhoven, the Netherlands*

[6]*Central Facility for Electron Microscopy (GFE), RWTH Aachen University, Ahornstr. 55, 52074 Aachen, Germany*

[7]*Ernst Ruska Centre for Microscopy and Spectroscopy with Electrons ER-C, Forschungszentrum Jülich GmbH, Jülich 52428, Germany*

[8]ROSI Freigeist Junior group, Soft Matter Physics and Optoelectronics, University of Potsdam, Potsdam, 14476, Germany

[9]Institute of Electronic Devices, University of Wuppertal, Rainer-Gruenter-Str. 21, 42119 Wuppertal, Germany

[10]Wuppertal Center for Smart Materials & Systems (CM@S), University of Wuppertal, Rainer-Gruenter-Str. 21, 42119 Wuppertal, Germany

[11]Institute of Biochemistry and Molecular Biology, Uniklinik RWTH Aachen, Pauwelsstrasse 30, Aachen, Germany

[12]Low-dimensional Semiconductors Lab, Department of Metallurgical and Materials Engineering, Indian Institute of Technology Madras, Chennai 600 036, India

*Corresponding Authors: mohammadi@amo.de; max.lemme@eld.rwth-aachen.de

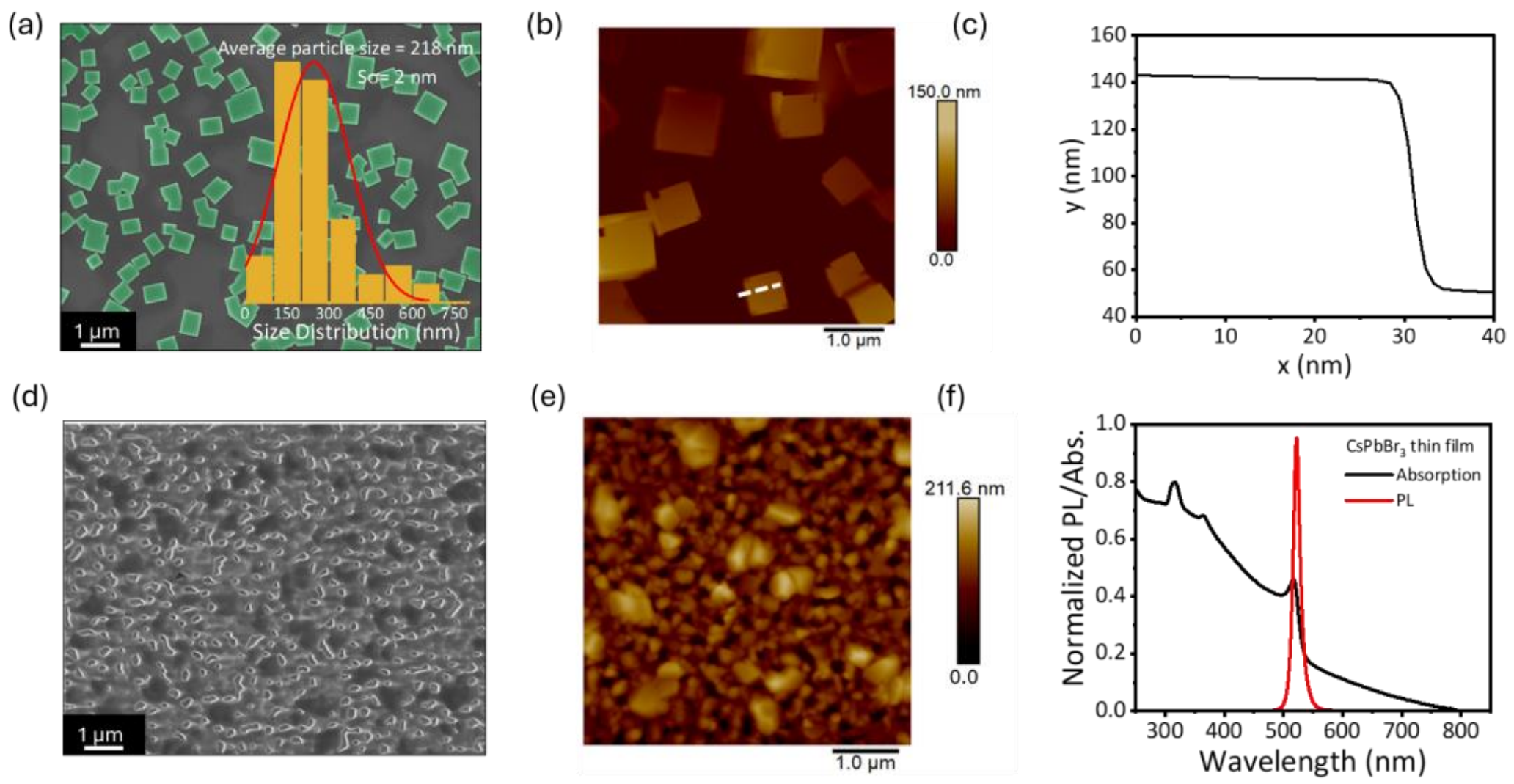

**Figure S1**. Morphology and Optical Properties of $CsPbBr_3$@TFA Nanoplatelets and reference $CsPbBr_3$. (a) Top-view SEM image of $CsPbBr_3$@TFA nanoplatelets. Inset: Size distribution statistics for nanoplatelets. (b) AFM topography of $CsPbBr_3$@TFA nanoplatelets. (c) AFM high-profile plot of $CsPbBr_3$@TFA nanoplatelets. (d, e) SEM image and AFM topography of a $CsPbBr_3$ reference sample. (f) UV-absorption and PL spectra of a $CsPbBr_3$ reference sample.

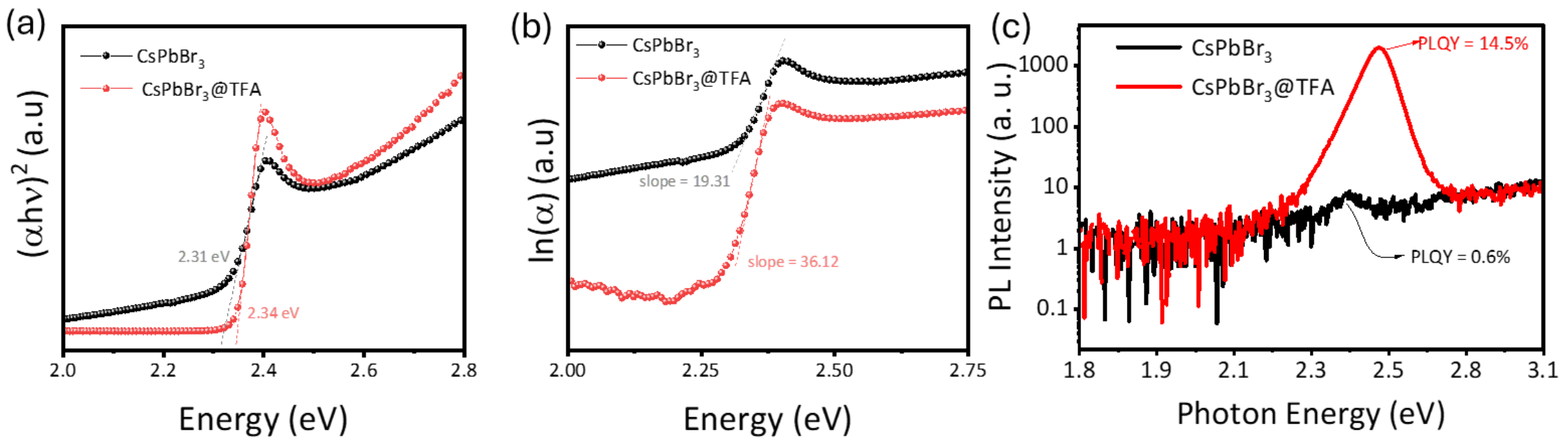

**Figure S2:** Optical absorption edge analysis of the $CsPbBr_3$ reference sample and $CsPbBr_3$@TFA. (a) Tauc plots used to determine the optical bandgap. (b) Urbach-tail analysis of the absorption edge. The steeper slope observed for $CsPbBr_3$@TFA indicates reduced band-tail broadening and a lower degree of structural disorder, consistent with a cleaner optical edge after TFA-mediated passivation. (c) PLQY measurements of the $CsPbBr_3$ reference sample and $CsPbBr_3$@TFA.

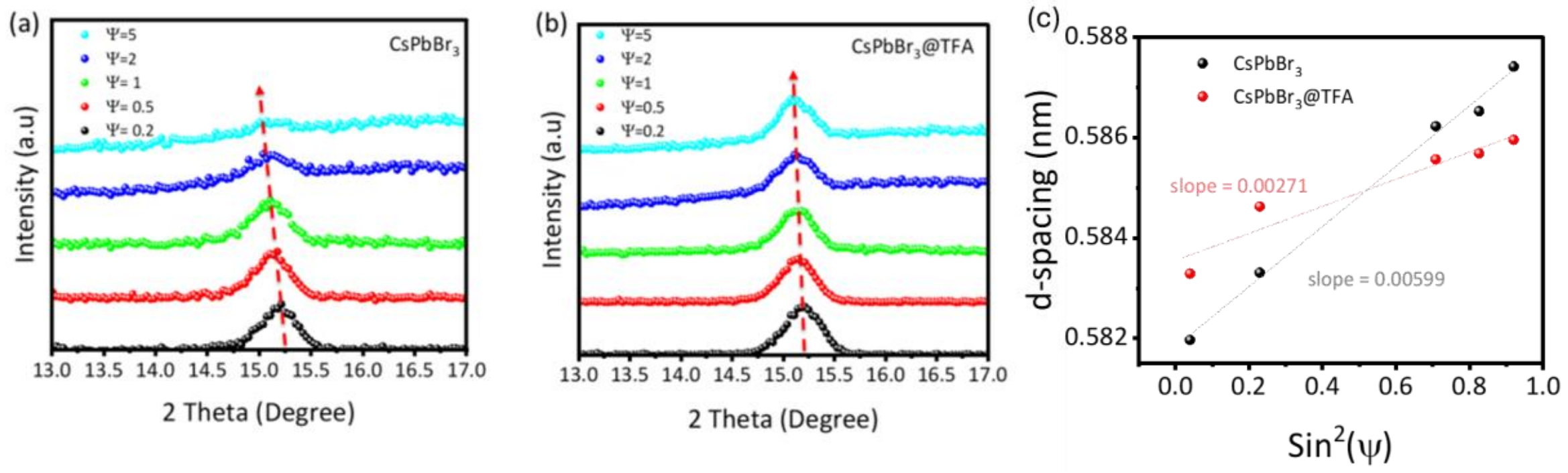

**Figure S3**. GIXRD analysis of the (001) plane for the perovskite samples at various tilt angles. (a) $CsPbBr_3$ reference sample, where diffraction peaks are shifted to the left, indicating residual strain in the film. (b) $CsPbBr_3$@TFA nanoplatelets, where the peak shift is significantly reduced, suggesting that $TFA^-$ effectively relieves strain. (c) Linear relationship between lattice spacing *d* and $\sin^2(\Psi)$ obtained from angle-dependent GIXRD measurements of the (001) plane for the reference sample and $CsPbBr_3$@TFA.

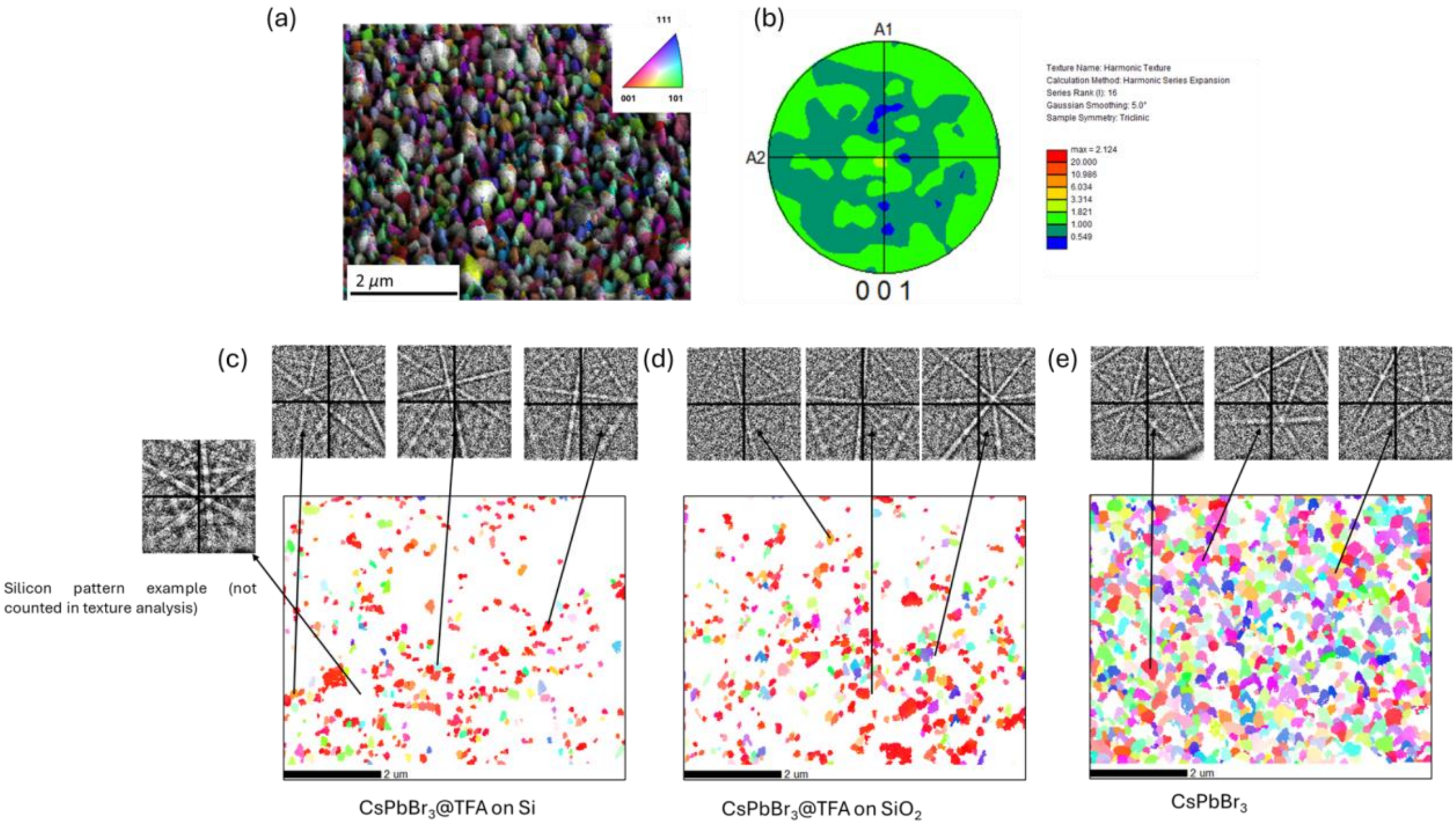

**Figure S4**. EBSD analysis of reference $CsPbBr_3$ and $CsPbBr_3$@TFA. (a) Inverse pole figure (IPF) map obtained by EBSD of the reference $CsPbBr_3$ deposited on a Si/$SiO_2$ substrate, showing grain orientations relative to the sample surface. (b) Corresponding [001] pole figure illustrating the crystallographic texture distribution. (c-e) Representative EBSD Kikuchi patterns and corresponding IPF/orientation overlays, extracted from the mapped regions for $CsPbBr_3$@TFA on Si, $CsPbBr_3$@TFA on $SiO_2$, and reference $CsPbBr_3$, illustrating pattern quality used for indexing. Bottom: the three datasets (scale bars: 2 μm).

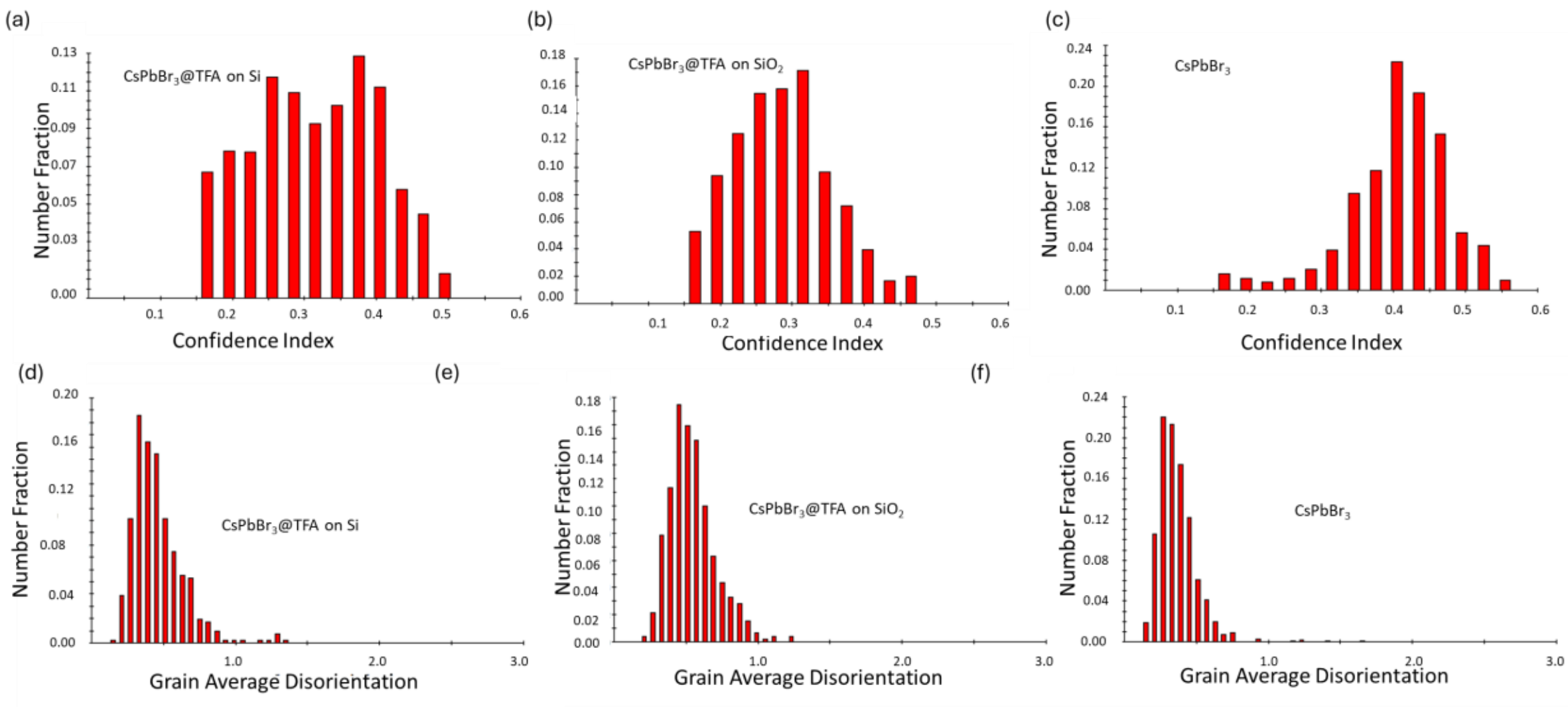

**Figure S5** EBSD indexing confidence and grain-resolved misorientation statistics. (a-c) Confidence index (CI) histograms for $CsPbBr_3$@TFA on Si, $CsPbBr_3$@TFA on $SiO_2$, and reference $CsPbBr_3$, respectively. (d-f) Corresponding grain-average disorientation/misorientation histograms for the same datasets. All statistics are computed from the data retained for texture evaluation after applying the CI mask (CI > 0.15).

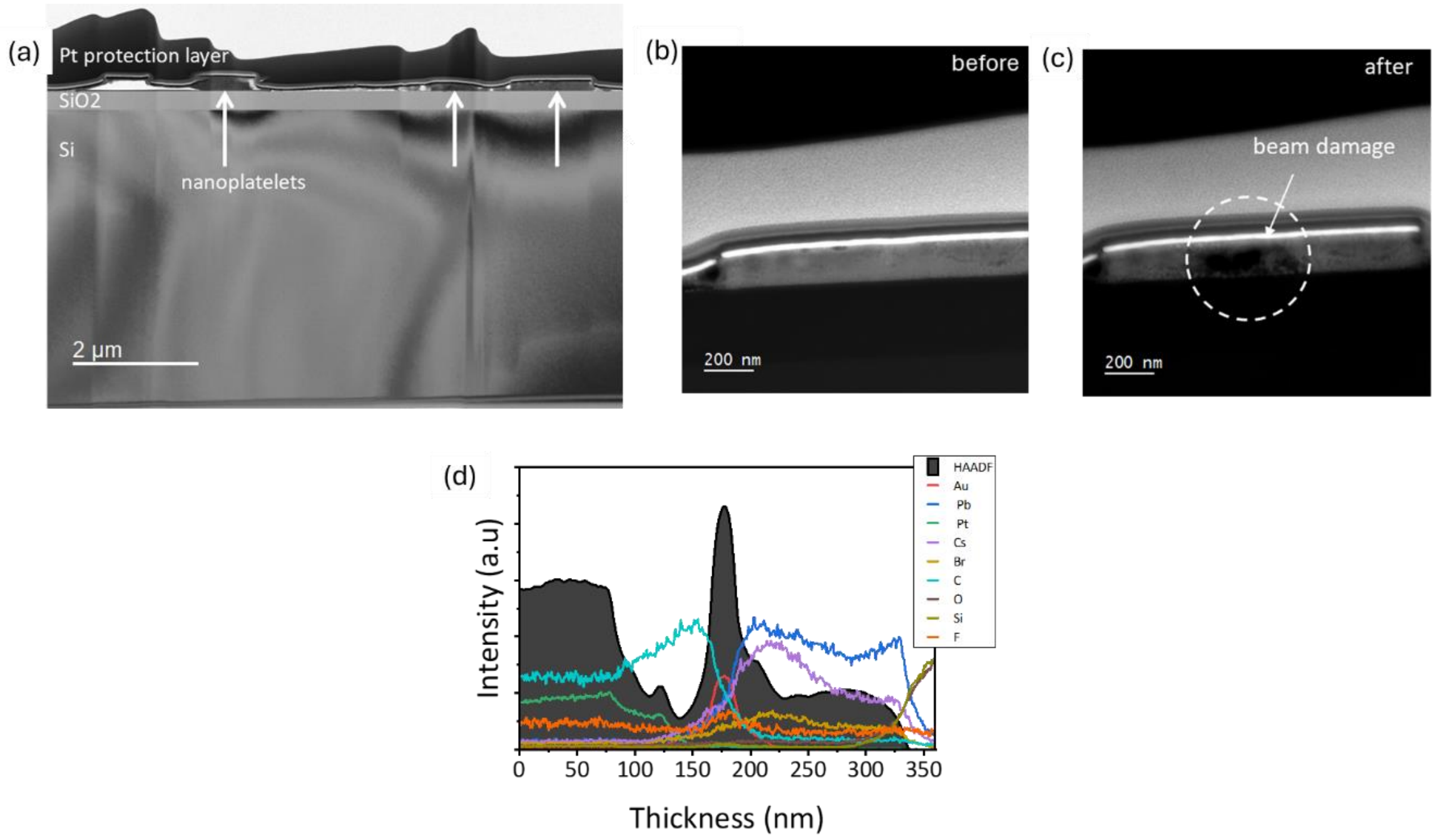

**Figure S6.** (a) Cross-sectional TEM lamella showing discrete nanoplatelets on the substrate. Effect of electron beam exposure on $CsPbBr_3$@TFA nanoplatelets. Cross-sectional STEM images of a nanoplatelet region, (b) before and (c) after electron-beam exposure, respectively. The sensitivity of the perovskite nanoplatelets to high-energy electron beams (200 kV) limits the acquisition of reliable crystalline phase information from the surface. (d) Line profile along the arrow in HAADF averaged within the box (shown in Figure 3d). The F peak is rather low and contains considerable background intensity.

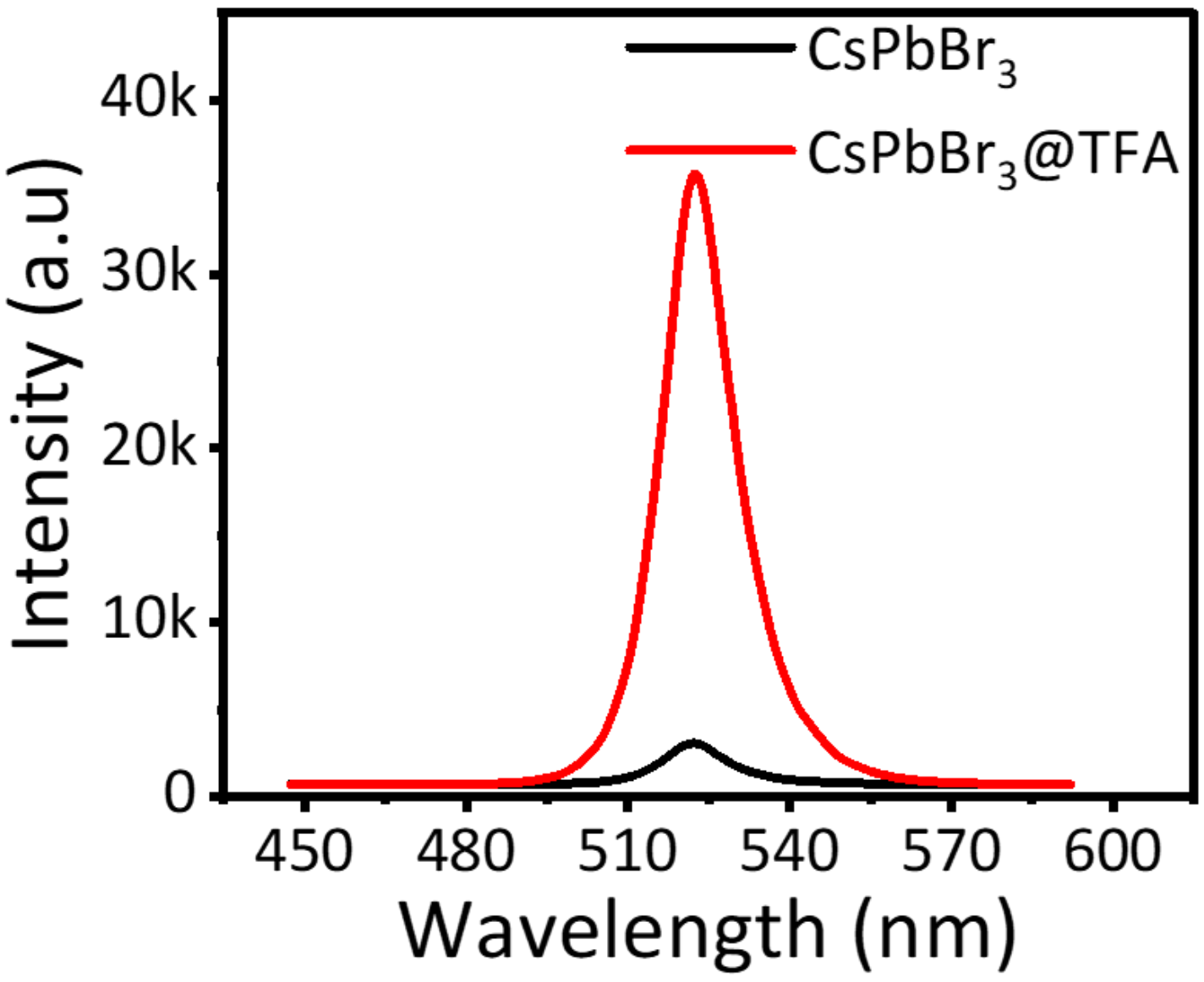

**Figure S7**. PL spectra showing the average PL intensity from $CsPbBr_3$@TFA nanoplatelets and $CsPbBr_3$ reference sample.

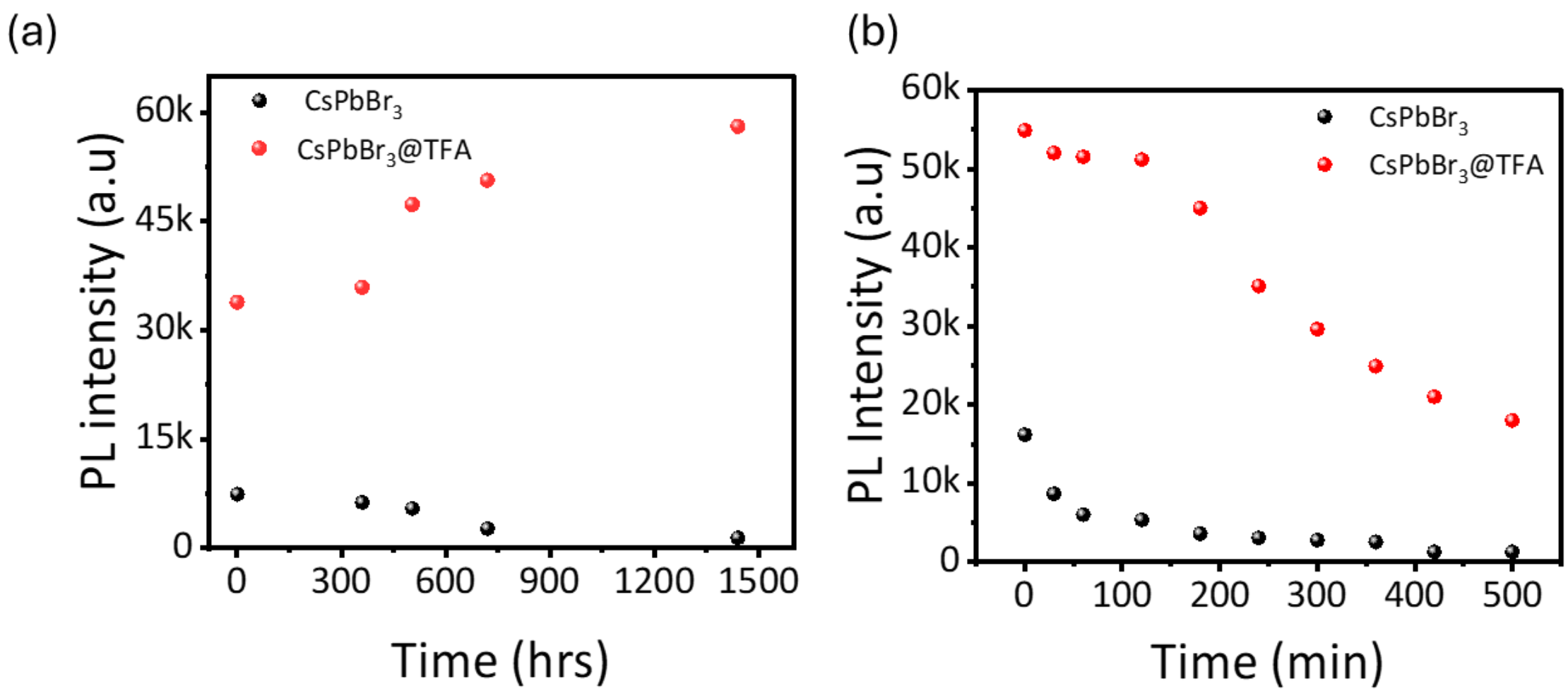

**Figure S8**. Optical emission stability of the $CsPbBr_3$@TFA nanoplatelets and $CsPbBr_3$ reference samples: (a) PL intensity as a function of storage time under ambient conditions. The samples were stored outside the glovebox at 46.1% relative humidity and 22 °C. (b) PL intensity under continuous-wave laser excitation with an irradiance of $2.83 \times 10^3$ W $cm^{-2}$ under ambient conditions.

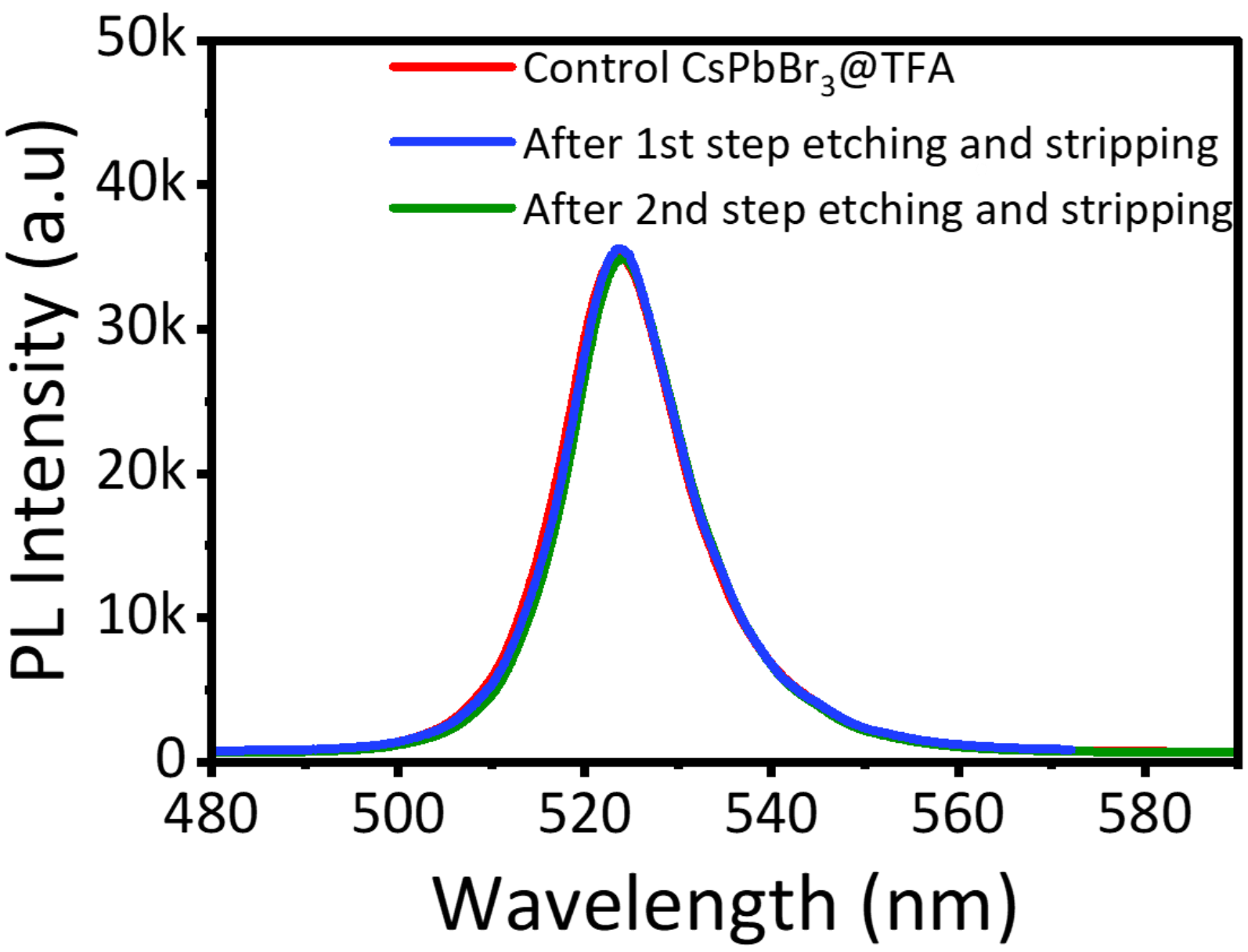

**Figure S9**. Optical emission stability of $CsPbBr_3$@TFA nanoplatelets during the photolithography process.

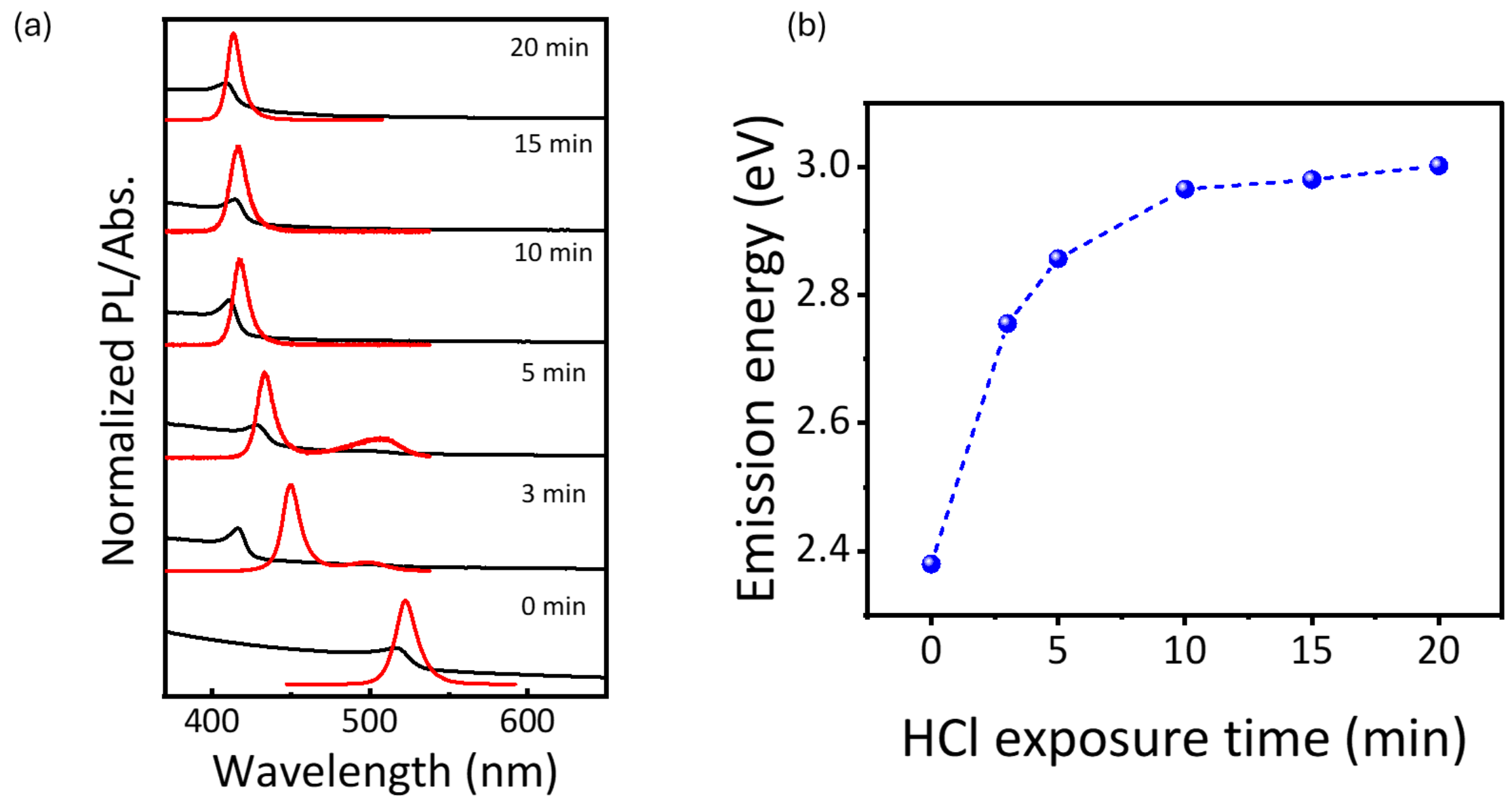

**Figure S10**. Time-dependent HCl-vapor exposure controls the emission of nanoplatelets. (a) Normalized PL and Absorption spectra recorded after different exposure times. (b) Dominant PL peak energy as a function of HCl exposure time.

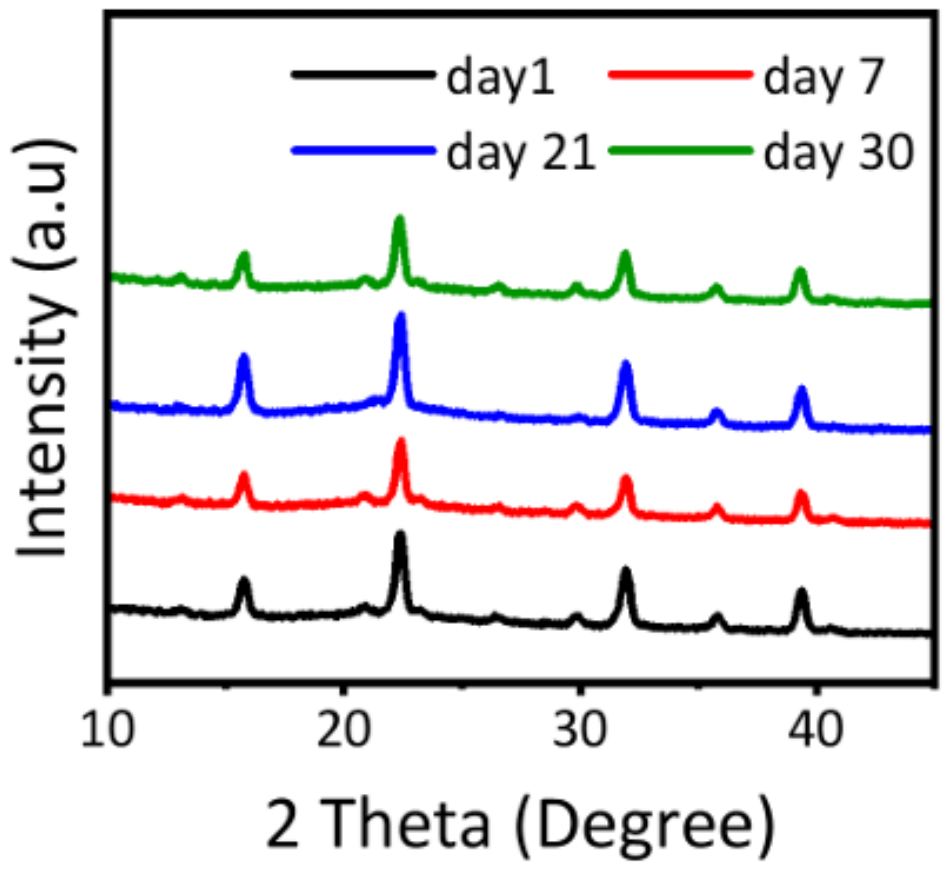

**Figure S11.** XRD patterns of the $CsPbCl_3$ nanoplatelets measured over a period of 30 days. The humidity was 46.1%, and the temperature was 22°C during the stability tests.

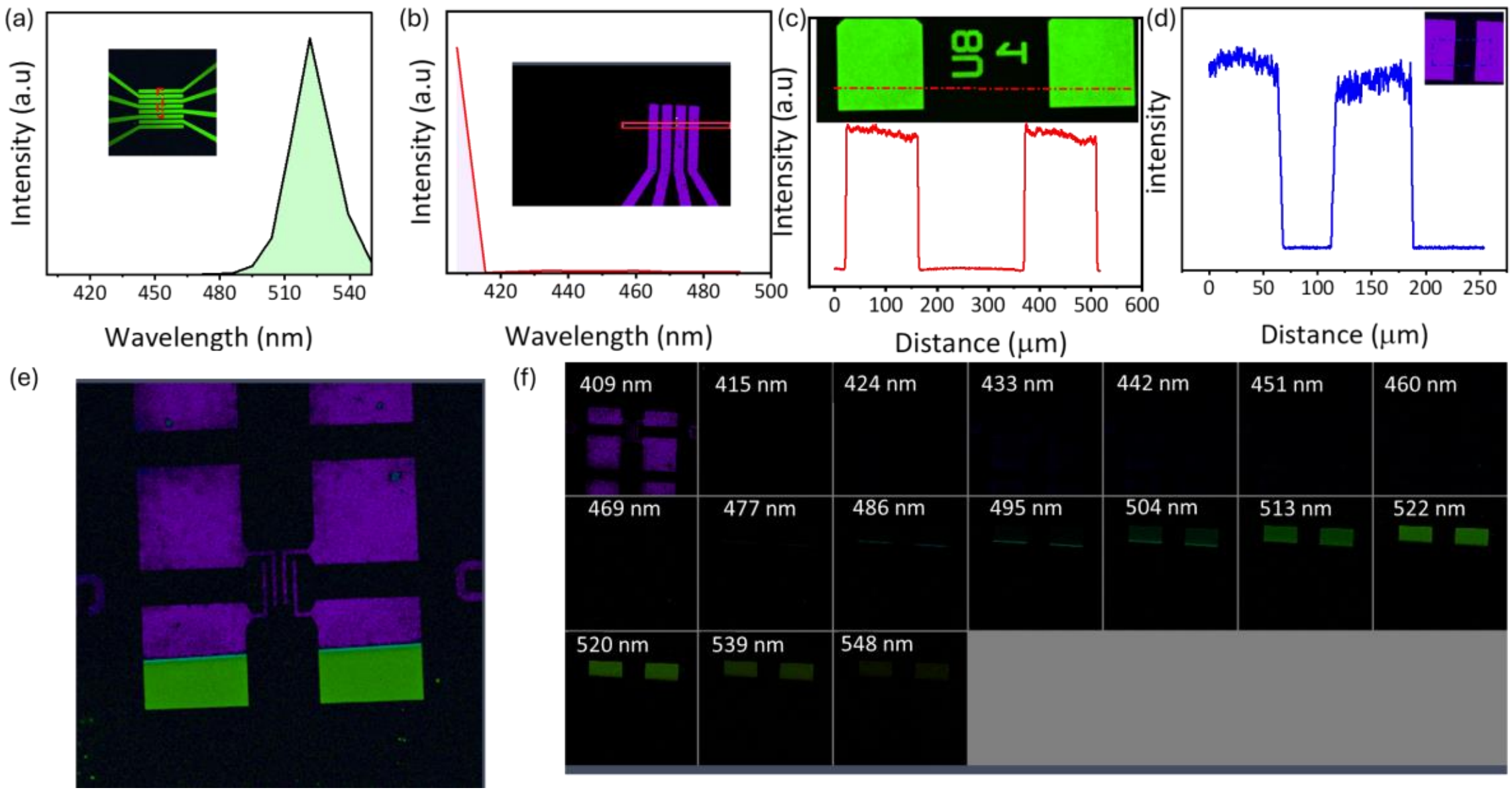

**Figure S12**. Confocal imaging of patterned nanoplatelets on a single chip. Emission spectrum acquired from representative: (a) masked regions, and (b) unmasked regions exposed to HCl vapors. Emission intensity as a function of distance shows the uniformity of perovskite for (c) masked regions, and (d) unmasked regions. (e) Confocal image of the patterned chip showing the coexistence of green-emitting protected $CsPbBr_3$@TFA regions and violet-emitting CsPbCl3@TFA regions. (f) Dimension annotation for several wavelengths from 409 nm to 548 nm, supporting distinct emission from the converted and unconverted regions.

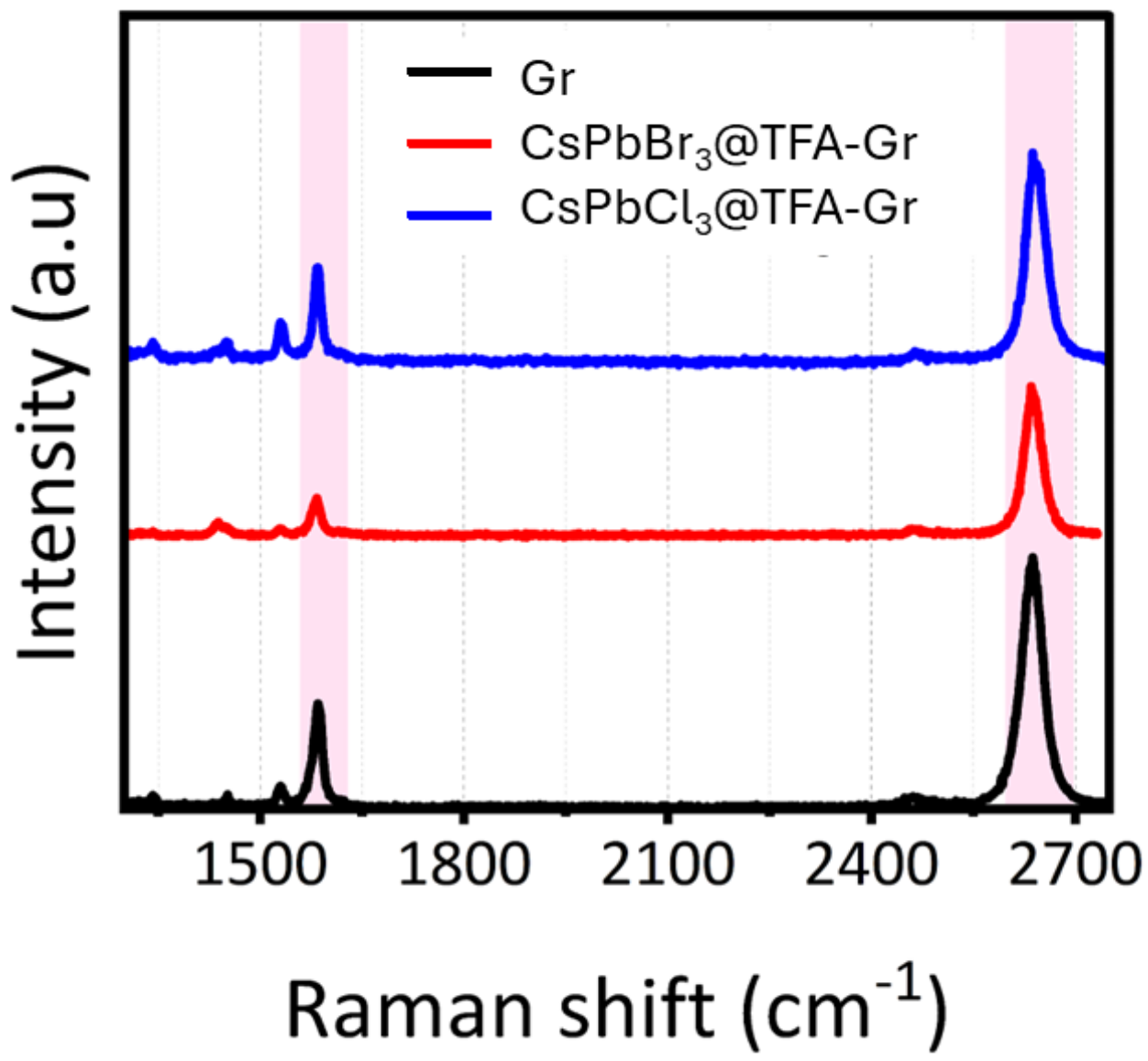

**Figure S13**. Raman Spectra of graphene, $CsPbbr_3$@TFA-graphene and $CsPbCl_3$@TFA -graphene.

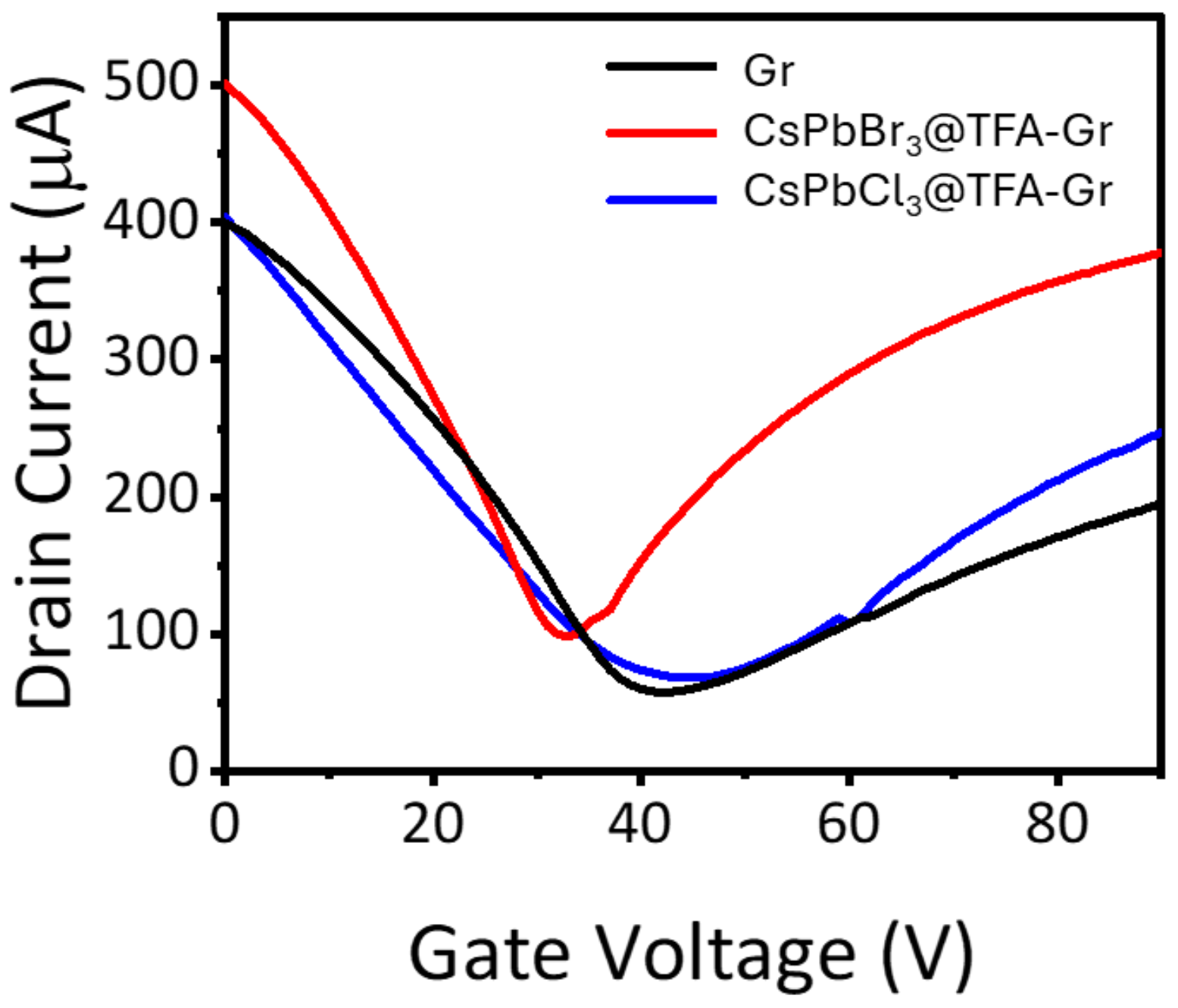

**Figure S14**. Transfer characteristics of GFET, comparing bare GFET and perovskite-integrated devices. The shift in the CNP and the change in minimum current reflect interfacial charge modulation induced by the nanoplatelets.

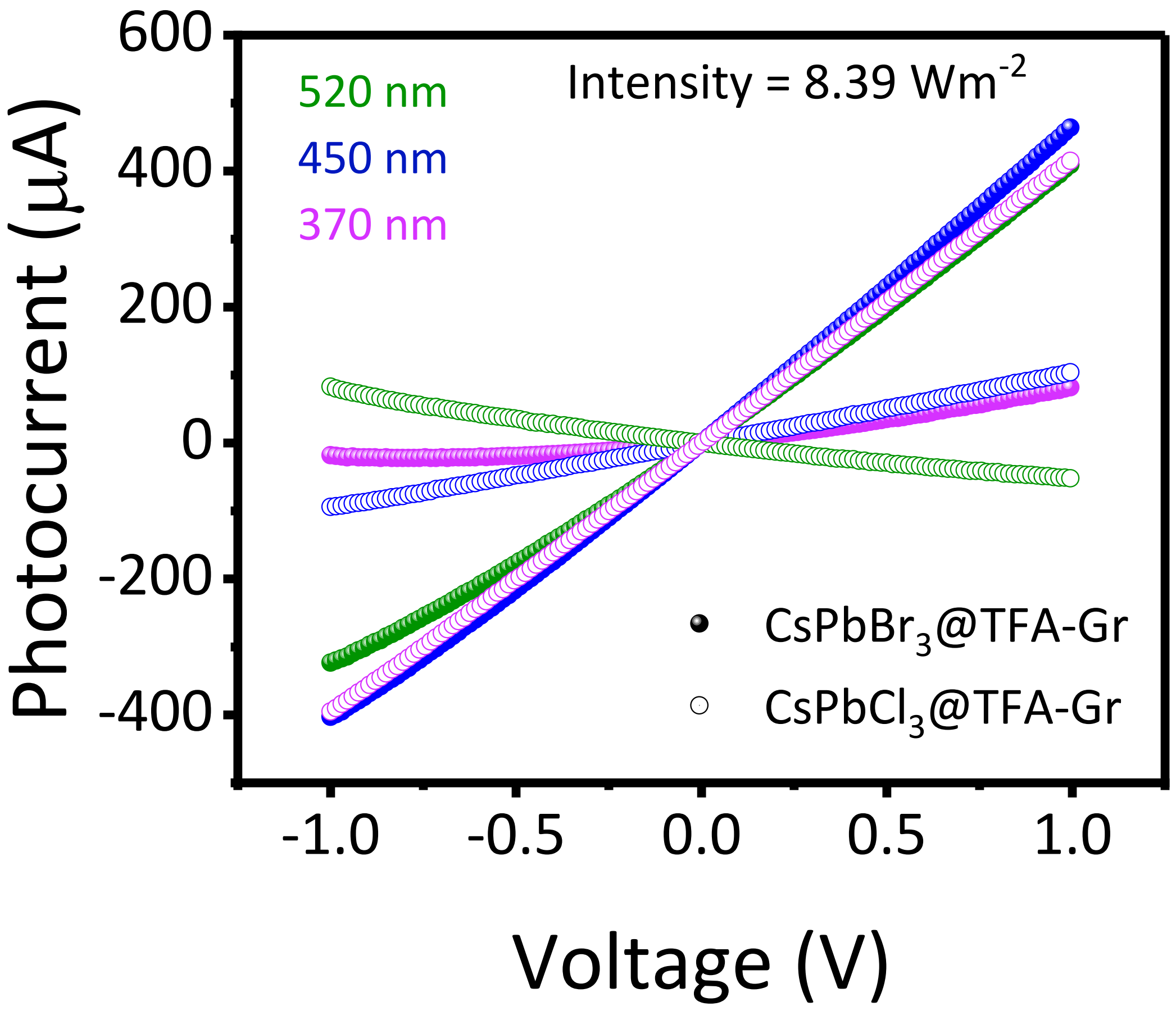

**Figure S15**. Photocurrent ($I_{ph}$= $I_{light}$-$I_{dark}$) of $CsPbBr_3$@TFA-graphene and $CsPbCl_3$@TFA-graphene hybrid transistor.

**Supplementary Note 1**

The TRPL decays were fitted using a bi-exponential function:[1]

$$I(t) = A_1 e^{-t/\tau_1} + A_2 e^{-t/\tau_2} \qquad (1)$$

where $A_1$, $A_2$ are amplitudes (counts) and $\tau_1$, $\tau_2$ are the corresponding decay constants.

**Table S1.** Bi-exponential fitting parameters of the TRPL decay curves for $CsPbBr_3$ and $CsPbBr_3$@TFA.

| Sample | A1 (counts) | $\tau_1$ (ns) | A2 (counts) | $\tau_2$ (ns) |
|---|---|---|---|---|
| $CsPbBr_3$ | 74405.82 | 1.43 | 1629.74 | 7.59 |
| $CsPbBr_3$@TFA | 39271.0 | 10.3 | 28540.0 | 33.24 |

To compare decays using a single metric, the intensity-weighted average lifetime ($\tau_{avg}$) was calculated as:[1]

$$\tau_{\mathrm{avg}} = \frac{A_1\tau_1^2 + A_2\tau_2^2}{A_1\tau_1 + A_2\tau_2} \quad (2)$$

The radiative and non-radiative recombination rates were estimated by combining the measured absolute PLQY with $\tau_{avg}$, using a simplified effective-lifetime model: [2]

$$\mathrm{PLQY} = \frac{k_r}{k_r + k_{nr}}, \tau_{\mathrm{avg}} = \frac{1}{k_r + k_{nr}} \quad (3)$$

which yields:

$$k_r = \frac{\mathrm{PLQY}}{\tau_{\mathrm{avg}}}, k_{nr} = \frac{1 - \mathrm{PLQY}}{\tau_{\mathrm{avg}}} \quad (4)$$

These rates provide a comparative measure of recombination behavior in the reference and TFA-derived samples but do not uniquely identify the underlying microscopic recombination pathways.

**Table S2.** Intensity-weighted average lifetime and extracted radiative/non-radiative recombination rates calculated using eq: 4

| Sample | PLQY (%) | $\tau_{avg}$ (ns) | ($k_r$) ($s^{-1}$) | ($k_{nr}$) ($s^{-1}$) |
|---|---|---|---|---|
| $CsPbBr_3$ | 0.6 | 2.07 | $2.90\times10^6$ | $4.80\times10^8$ |
| $CsPbBr_3$@TFA | 14.5 | 26.38 | $5.50\times10^6$ | $3.24\times10^7$ |